\documentclass{aa}

\usepackage{txfonts}
\usepackage[none]{hyphenat}
\usepackage{siunitx}
\usepackage[thinc]{esdiff}
\usepackage{mathtools}
\usepackage{amsmath}
\usepackage{graphicx}
\usepackage[thinc]{esdiff}
\usepackage{diagbox}
\usepackage{soul}
\usepackage{placeins}

\usepackage{times}

\usepackage[T1]{fontenc}
\usepackage{CJKutf8}
\usepackage{aecompl}
\usepackage{calligra}
\DeclareMathAlphabet{\mathcalligra}{T1}{calligra}{m}{n}
\DeclareFontShape{T1}{calligra}{m}{n}{<->s*[2.2]callig15}{}

\usepackage{xcolor}
\definecolor{darkblue}{RGB}{12,94,176}
\usepackage{hyperref}
\hypersetup{
    pdfnewwindow=true,     
    colorlinks=true,       
    linkcolor=darkblue,    
    citecolor=darkblue,    
    filecolor=darkblue,    
    urlcolor=blue          
}

\def\apjl{ApJL}               
\def\apjs{ApJS}               
\def\aap{A\&A}                

\usepackage{color}

\usepackage[acronym,nohypertypes={acronym}]{glossaries}
\newacronym{smbh}{SMBH}{Super Massive Black Hole}
\newacronym{bh}{BH}{Black Hole}
\newacronym{ce}{CE}{Common Envelope}
\newacronym{cbd}{CBD}{CircumBinary Disk}

\newcommand{\orcid}[1]{\unskip\protect\href{https://orcid.org/#1}{\protect\includegraphics[width=8pt,clip]{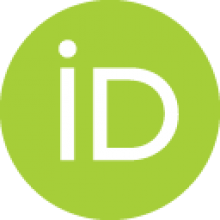}}}

\DeclareSIUnit\parsec{pc}
\DeclareSIUnit\cMpc{cMpc}
\DeclareSIUnit\year{yr}
\DeclareSIUnit\Zsun{Z_{\odot}}
\DeclareSIUnit\Msun{M_{\odot}}
\DeclareSIUnit\Rsun{R_{\odot}}
\DeclareSIUnit\Lsun{L_{\odot}}
\DeclareSIUnit\erg{erg}
\DeclareSIUnit\eV{eV}
\DeclareSIUnit\au{AU}

\newcommand{\GG}[1]{}

\newcommand{\tableref}[2]{\footnotesize{\textsuperscript{[#1]} #2}}

\graphicspath{{./}{figures/}}

\begin{document}
\begin{CJK*}{UTF8}{gbsn}

\title{Long-term evolution of binary orbits induced by circumbinary disks}

\author{Ruggero Valli\thanks{ruvalli@mpa-garching.mpg.de} \fnmsep \inst{1} \orcid{0000-0003-3456-3349}
    \and
    Christopher Tiede \inst{2} \orcid{0000-0002-3820-2404}
    \and
    Alejandro Vigna-G\'omez \inst{1} \orcid{0000-0003-1817-3586}
    \and
    Jorge Cuadra \inst{3,4} \orcid{0000-0003-1965-3346}
    \and
    Magdalena Siwek \inst{5} \orcid{0000-0002-1530-9778}
    \and\\
    Jing-Ze Ma (马竟泽) \inst{1} \orcid{0000-0002-9911-8767}
    \and
    Daniel J. D'Orazio \inst{2} \orcid{0000-0002-1271-6247}
    \and
    Jonathan Zrake \inst{6} \orcid{0000-0002-1895-6516}
    \and
    Selma E. de Mink \inst{1,7} \orcid{0000-0001-9336-2825}
}

\institute{Max-Planck-Institut f{\"u}r Astrophysik, Karl-Schwarzschild-Straße 1, 85741 Garching, Germany
    \and
    Niels Bohr International Academy, Niels Bohr Institute, Blegdamsvej 17, 2100 Copenhagen, Denmark
    \and
    Departamento de Ciencias, Facultad de Artes Liberales, Universidad Adolfo Ib{\'a}{\~n}ez, Av. Padre Hurtado 750, Vi{\~n}a del Mar, Chile
    \and
    N{\'u}cleo Milenio de Formaci{\'o}n Planetaria (NPF), Chile
    \and
    Center for Astrophysics, Harvard University, Cambridge, MA 02138, USA
    \and
    Department of Physics and Astronomy, Clemson University, Clemson, SC 29634, USA
    \and
    Anton Pannekoek Institute for Astronomy and GRAPPA, University of Amsterdam, NL-1090 GE Amsterdam, The Netherlands
}

\abstract{
Circumbinary disks are found in a variety of astrophysical scenarios, spanning binary star formation to accreting supermassive black hole binaries. Depending on the characteristics of the system, the interaction with a circumbinary disk can either damp or excite the binary's eccentricity and can also widen or shrink the orbit.

To predict the outcome of the long-term disk-binary interaction, we present a new formalism based on the results of recent suites of hydrodynamic simulations, which resolve the complex geometry of the gas in the vicinity of the binary and fully account for the gravitational and accretion forces.

We released a python package, \texttt{spindler}, that implements our model. We show that---under the assumed thin disk model with a fixed thickness and viscosity prescription---accretion onto the binary depletes the disk mass before inducing a significant change in the orbital separation or the mass ratio, unless the mass reservoir feeding the disk is comparable to the mass of the binary. This finding implies that, in most scenarios, an interaction with a circumbinary disk is not an efficient mechanism to shrink the orbit of the binary. However, the interaction can excite the eccentricity up to an equilibrium value, and induce a statistical correlation between the mass ratio and eccentricity, as long as the mass of the disk is at least a few percent of the mass of the binary.

We consider the applicability of our model to a variety of astrophysical scenarios: during star formation, in evolved stellar binaries, triples, and in supermassive black hole binaries. We discuss the theoretical and observational implications of our predictions.}


\maketitle
\end{CJK*}

\section{Introduction} \label{sec:intro}


Circumbinary disks are pervasive in the Universe. They exist across astrophysical scales, ranging from supermassive black hole (SMBH) binaries \citep{begelman_1980, haiman_population_2009, roedig_limiting_2011, franchini_circumbinary_2021, dorazio_observational_2023} to stellar systems. For stellar binaries, such disks are present in the early stages during star formation \citep{dutrey_images_1994, mathieu_classical_1997, takakuwa_angular_2014, lacour_m-dwarf_2016, tofflemire_pulsed_2017, long_2021} as well as during the late stages for post-asymptotic giant branch (AGB) stars \citep{deroo_amber_2007, van_winckel_post-agb_2009,  bujarrabal_extended_2013-1, gallardo_cava_keplerian_2021} and in post-common envelope systems \citep{kashi_circumbinary_2011, reichardt_extending_2019, ropke_simulations_2023, tuna_long-term_2023}. They may further occur in triple systems when the outer star donates  mass to the inner system \citep{di_stefano_mass_2020, hamers_statistical_2022, dorozsmai_stellar_2024}. One of the primary goals in the study of circumbinary disks is to understand how the interaction with the disk influences the orbital evolution of the binary. The intricate interplay between the time-varying binary potential, the disk's morphological response, and the resulting forces on the binary orbit can lead to a wide array of outcomes including secular changes in the binary orbital elements, a variation in the mass accretion, and possibly even mergers.

A substantial body of research, spanning from early analytical and numerical investigations \citep{artymowicz_role_1983, artymowicz_1991, lubow_young_1996, pringle_properties_1991} to more recent hydrodynamic simulations \citep[e.g.,][]{macfadyen_eccentric_2008, cuadra_2009, shi_three-dimensional_2012, dorazio_accretion_2013, munoz_hydrodynamics_2019}, reveals the sensitivity of these outcomes to the parameters of the system. The current understanding delineates a complicated landscape within a parameter space that remained, until recently, relatively uncharted. The evolution of the orbital parameters is a nontrivial function of mass ratio \citep{ragusa_evolution_2020, duffell_circumbinary_2020, siwek_preferential_2023, siwek_orbital_2023, dittmann_evolution_2024} and eccentricity \citep{miranda_viscous_2017,zrake_equilibrium_2021, siwek_orbital_2023}; and can also be influenced by the disk aspect ratio \citep{young_binary_2015,ragusa_suppression_2016, tiede_gas-driven_2020, dittmann_survey_2022, dittmann_evolution_2024}, viscosity \citep{duffell_circumbinary_2020, dittmann_survey_2022, dittmann_evolution_2024, turpin_orbital_2024}, disk-orbit inclination \citep{moody_hydrodynamic_2019, smallwood_2022, tiede_eccentric_2024, martin_mergers_2024, bourne_dynamics_2023}, and cooling rate \citep{wang_role_2023, wang_hydrodynamical_2023}.

Because of these complex dependencies, fully characterizing binary-disk interactions requires extensive parameter exploration.
These studies are typically performed either in 3D smoothed particle hydrodynamics \citep[SPH,][]{lucy_numerical_1977, gingold_smoothed_1977} where the orbit of the binary is integrated self-consistently from the forces of the gas particles, or with grid-based simulations where it is typical to fix a Keplerian orbit and compute the change in orbital elements in post-processing from the gas forces.
These studies can be computationally demanding because of the high spatial resolutions and long integration times required to achieve converged results, limiting the possibility of self-consistently following the long-term evolution of the system. 

Due to the resources required by hydrodynamic simulations, previous works interested in modeling the long-term effect of the disk on the orbit preferred to rely on analytical prescriptions. In some cases, the employed prescriptions are based on analytical disk models, which did not take into account the complex structure and dynamics of the gas inside the central cavity \citep{rafikov_eccentricity_2016, siwek_effect_2020,izzard_circumbinary_2023, wei_evolution_2023} and sometimes neglected accretion forces \citep{dermine_eccentricity-pumping_2013, vos_testing_2015}. Other works chose to arbitrarily fix the speed of orbital shrinking \citep{tokovinin_formation_2020, tuna_long-term_2023} or neglected the angular momentum exchanged with the disk \citep{di_stefano_mass_2020, hamers_statistical_2022, dorozsmai_stellar_2024}.

In this work, we propose a new approach to compute the long-term interaction of the binary with the circumbinary disk. Our approach is based on the results of high-resolution hydrodynamic simulations, which take into account both the gravitational and accretion forces.
Specifically, we integrated across the parameter space the derivatives obtained from hydrodynamic simulations in \citet{siwek_preferential_2023, siwek_orbital_2023}, which we then compared to the results of \citet{zrake_equilibrium_2021} and \citet{dorazio_orbital_2021}.
In Sect.~\ref{sec:methods} we describe the physical assumptions behind the models and the numerical strategy employed to compute the long-term evolution. In Sect.~\ref{sec:results} we show the trajectories of individual systems in the eccentricity--mass ratio--semi-major axis parameter space, and identify possible observational signatures for a population of systems. Section~\ref{sec:comparing_results} is dedicated to the comparison with the models from \citet{zrake_equilibrium_2021} and \citet{dorazio_orbital_2021}, wherein we highlight key differences and similarities. In Sect.~\ref{sec:implications} we discuss the implications and applicability of our results to different astrophysical scenarios. Lastly, we summarize our findings in Sect.~\ref{sec:conclusion}.

\section{Methods} \label{sec:methods}
We aim to model the long-term orbital evolution of a population of binaries over the lifecycle of disk-mediated accretion phases.
Employing hydrodynamic simulations directly for this purpose would be prohibitively expensive. However, we can use the results from fixed-parameter simulations to build a model of the long-term effects of the disk-binary interaction on the orbit. Our methodology involves three steps.
\begin{enumerate}
    \vspace{-5pt}
    \item  We collect the results of hydrodynamic simulations of circumbinary disks available in the existing literature. 
    From these simulations, we obtain the derivatives describing the rates at which the separation $a$,  eccentricity $e$, and mass ratio $q$ change as a function of accreted mass (Section~\ref{sec:hydro_description} and \ref{sec:hydro_comparison}, the values used are reported in Appendix~\ref{app:tables_of_derivatives}).
    \item We then interpolate these derivatives with a continuous function between the points of the parameter space that were covered by hydrodynamic simulations (Section~\ref{sec:interpolation}). 
    \item Finally, we use the function obtained in step 2 to numerically solve for the long-term evolution of a system after specifying its initial parameters $a_0$, $e_0$, and $q_0$. (Section~\ref{sec:integration}).
\end{enumerate}

We release \texttt{spindler}, a python package that performs these calculations, available as a Zenodo release\footnote{\href{https://doi.org/10.5281/zenodo.10529060}{doi.org/10.5281/zenodo.10529060}} and as a github repository\footnote{\href{https://github.com/ruggero-valli/spindler}{github.com/ruggero-valli/spindler}}.
The code is modular, allowing for future extensions to incorporate findings from additional hydrodynamic simulations, for example, to explore different viscosities or disk aspect ratios. We also release a c-based version of  \texttt{spindler}\footnote{\href{https://github.com/ruggero-valli/spindler-c}{github.com/ruggero-valli/spindler-c}}, for better compatibility with c and Fortran based codes.
The interpolation routines of the c version rely on the \texttt{librinterpolate} library\footnote{\href{https://gitlab.com/rob.izzard/librinterpolate}{gitlab.com/rob.izzard/librinterpolate}} by Robert Izzard.

\subsection{Fiducial suite of hydrodynamics simulations}\label{sec:hydro_description}

\defcitealias{siwek_preferential_2023}{S23a}
\defcitealias{siwek_orbital_2023}{S23b}
\defcitealias{zrake_equilibrium_2021}{Z21}
\defcitealias{dorazio_orbital_2021}{DD21}

As a fiducial model, we employ the results of a suite of 2D hydrodynamic simulations carried out and analyzed by \citet[][hereafter \citetalias{siwek_preferential_2023} and \citetalias{siwek_orbital_2023}]{siwek_preferential_2023, siwek_orbital_2023}  using the moving-mesh code \texttt{AREPO} \citep{springel_e_2010, pakmor_improving_2016, munoz_multidimensional_2013}. Here, we summarize the key underlying assumptions.

The simulations represent a finite gas disk accreting onto a central binary on a coplanar prograde orbit of semi-major axis $a$ and eccentricity $e$. 
They are performed in 2D, assuming vertical hydrostatic equilibrium with a locally isothermal equation of state.
The two objects composing the binary are modeled as sink particles of mass $M_1$ and $M_2$ (where $M_1>M_2$) and total mass $M \equiv M_1+M_2$. These simulations are designed to capture the regime where the objects' radii are substantially smaller than their separation. Mass is removed from the system from a region within a radius of $0.03a$ from the sink particles. 
The sink radius is larger than the objects' horizon or surface radius, but is determined as small enough to not influence the solutions. 

The disk aspect ratio is set to $h/r=0.1$, where $h$ is the thickness of the disk and $r$ is the radial coordinate. Viscosity is treated with the $\alpha$-disk prescription \citep{shakura_black_1973}, where Reynolds stresses from unresolved turbulence are characterized by the dimensionless parameter $\alpha=0.1$. The gravitational potential of the binary is softened with a softening length of $0.025a$, and the mass of the disk is taken to be much smaller than the mass of the binary so that the self-gravity of the disk can be ignored.

Under the stated assumptions, the system of equations becomes scale-free, leaving only two free parameters: the binary mass ratio $q \equiv M_2/M_1 \leq 1$ and the eccentricity of the binary orbit $e$. \citetalias{siwek_preferential_2023} explored a rectangular grid of simulations considering ten different mass ratios within the range $0.1 \leq q \leq 1$, and five values for the orbital eccentricity, spanning from $e=0$ to $e=0.8$.
\citetalias{siwek_orbital_2023} expanded the grid to nine different eccentricities, spanning the same eccentricity range.

The main quantities we are interested in for this work are the time derivatives of the orbital eccentricity  $\dot{e}$, separation $\dot{a}$, and mass ratio $\dot{q}$, since these will set the orbital evolution of the binary systems.
The binary is not evolved directly in the simulation, as the orbit is kept fixed. Instead, the interaction with the circumbinary disk is tracked by recording the rate of change of angular momentum, energy, and mass at each simulation time-step. From these quantities, the derivatives of the orbital parameters are computed and then averaged over hundreds of orbits, sufficient to remove the effect of the orbital variability, the variability associated with the lump precession \citep[which occurs on a timescale of about five orbits, ][]{macfadyen_eccentric_2008, shi_three-dimensional_2012, dorazio_accretion_2013, farris_binary_2014, dittmann_survey_2022}, and the cavity precession \citep[about 300 orbits,][]{munoz_eccmodes_2020, siwek_preferential_2023}.

For convenience, instead of referring to time derivatives, we express the derivatives in terms of the variation of total binary mass:
\begin{equation} \label{eq:def_derivatives}
q' \equiv \diff{q}{\log{M}}, \quad
a' \equiv \diff{\log{a}}{\log{M}}, \quad
e' \equiv \diff{e}{\log{M}}.
\end{equation}
These expressions have the advantage of being invariant under the rescaling of the simulation's physical units of binary mass and gas surface density.

Our approach to compute the long-term orbital evolution is valid when in a quasi-equilibrium state, that is the limit where the parameters of the binary vary on a timescale much longer than the viscous relaxation time $t_{\rm visc}$ of the inner region of the disk, which dominates the forces on the binary. This condition is enforced in the simulations, but it depends on the astrophysical system of interest whether it is satisfied.

\subsection{Additional simulations}
\label{sec:hydro_comparison}

\begin{table*}[!ht]
    \centering
    \caption{Summary of the main parameters and results of the considered suites of hydrodynamic simulations.}
    \begin{tabular}{|c|ccc|}
    \hline
                 & \citet{siwek_preferential_2023, siwek_orbital_2023} & \citet{zrake_equilibrium_2021} & \citet{dorazio_orbital_2021} \\
                 & \citetalias{siwek_preferential_2023}, \citetalias{siwek_orbital_2023} & \citetalias{zrake_equilibrium_2021}  &
                 \citetalias{dorazio_orbital_2021} \\ \hline
                 
        Code &  \texttt{Arepo} & \texttt{Mara3} & \texttt{DISCO} \\ 
        Mesh type & Voronoi moving mesh & Cartesian grid & Cylindrical moving mesh \\ 
        Viscosity & $\alpha=0.1$ & $\alpha=0.1$ & $\nu= 10^{-3} a^2 \Omega$ \\ 
        Disk aspect ratio & 0.1 & 0.1 & 0.1 \\ 
        Gravity softening radius [$a$] & 0.025 & 0.02 & 0.05 \\ 
        Sink radius [$a$] & 0.03 & 0.02 & 0.1 \\
        Parameter space ($e$-$q$) & $0.1\leq q \leq 1$ & $q=1$ & $q=1$ \\        
         & $0 \leq e \leq 0.8$ & $0 \leq e \leq 0.8$ & $0 \leq e \leq 0.9$ \\        
        Exploration strategy & discrete sampling & discrete sampling & adiabatic sweeping \\ 
        Parameter combinations explored & $9 \times 10$ & 32 & $\approx 40$\textsuperscript{[a]} \\
        Number of orbits per combination & 10\,000 & 2000 & $\approx 444$\textsuperscript{[a]} \\
        Relaxation radius [$a$] & $\approx 21$\textsuperscript{[b]} & $\approx 7$ & $\approx 2$\textsuperscript{[a]} \\       
        \hline
        Circular orbit stability & unstable & stable & stable \\
        $e_{\rm eq}$ & $0.48^{+0.02}_{-0.08}$ & $0.443^{+0.007}_{-0.018}$ & $0.39^{+0.01}_{-0.01}$ \\
        $a'(e{=}e_{\rm{eq}})$ & -0.70 & -2.3 & 0 or -0.49\textsuperscript{[c]} \\ \hline
        \multicolumn{4}{l}{\footnotesize{\textsuperscript{[a]}estimated, given a single sweep 20\,000 orbits long and a resolution in eccentricity $\Delta e=0.02$}}\\
        \multicolumn{4}{l}{\footnotesize{\textsuperscript{[b]}\citetalias{siwek_preferential_2023} report 27 instead of 21, due to a typo in equation~7. The correct factor in front of the viscous timescale is $2/3$ instead of $4/9$.}}\\
        \multicolumn{4}{l}{\footnotesize{\textsuperscript{[c]}if considering oscillations around the equilibrium configuration.}}\\
        
    \end{tabular}

    \label{tab:hydro_comparison}
\end{table*}

To better understand the impact of model uncertainties, we compare our main results, based on  hydrodynamic simulations by \citetalias{siwek_orbital_2023}, with alternative simulations  by  \citet[][hereafter \citetalias{zrake_equilibrium_2021}]{zrake_equilibrium_2021} and \citet[][hereafter \citetalias{dorazio_orbital_2021}]{dorazio_orbital_2021}. These works modeled circumbinary accretion and the effect on the orbit under nearly identical assumptions.
The main difference from \citetalias{siwek_orbital_2023} is that these two studies focus exclusively on $q=1$ binaries, but with a far denser sampling in eccentricity.
Below we detail the other modeling variations, which are additionally summarized in Table~\ref{tab:hydro_comparison}.

Beyond the differences in parameter exploration, the primary distinction between the three works is the structure of the underlying simulation grid.
\citetalias{siwek_orbital_2023} uses the Voronoi moving mesh of \texttt{Arepo}, \citetalias{zrake_equilibrium_2021} employs the Cartesian grid-based code with static mesh-refinement  \texttt{Mara3} \citep{zrake_numerical_2012, tiede_gas-driven_2020}, and \citetalias{dorazio_orbital_2021} rely on the cylindrical moving mesh of \texttt{DISCO} \citep{duffell_disco_2016, duffell_circumbinary_2020}.
Each technique has its respective strengths. Moreover, compiling the results bolsters confidence in features on which the three agree. Conversely, our approach diminishes the likelihood of capturing dynamics that may be emergent from the underlying mesh structure.

The treatment of viscosity in the three studies is nearly identical.
\citetalias{siwek_orbital_2023} and \citetalias{zrake_equilibrium_2021} both employ an $\alpha$-disk prescription \citep{shakura_black_1973}, setting $\alpha=0.1$ (see the appendix of \citetalias{siwek_preferential_2023} on the implementation for a binary potential). \citetalias{dorazio_orbital_2021} assumes a constant kinematic viscosity $\nu = 10^{-3} a^2 \Omega$, where $\Omega=2 \pi/P_{\rm{orb}}$ is the orbital frequency and $P_{\rm{orb}}$ is the orbital period. The two prescriptions yield a comparable value of viscosity at $r=a$, for $h/r = 0.1$.

All three studies treat accretion onto the individual components of the binary with a sink particle at the location of the binary objects. 
The treatment of the sink particles can affect the morphology of the ``mini-disks'' formed around them, as well as the rate at which mass and angular momentum are imparted onto the binary.
All three studies employ the commonly used sink prescription for which the removed angular momentum is the one associated with the removed mass at the local fluid velocity (as opposed to ``torque-free'' or ``spinless'' sinks, \citealt{dempsey_sinks_2020}).
As such, all of these studies will possess some induced spin-component in the binary angular momentum (and this spin component would be real in the limit of a resolved stellar surface or black hole Innermost Stable Circular Orbit, ISCO). 
This could be relevant for some subsets of accreting binaries, and while the effects of this spin component have been documented for circular binaries \citep[e.g.,][]{munoz_hydrodynamics_2019, dittmann_sinks_2021}, they have not been examined in detail for eccentric orbits.
The sink radius is always taken as a small fraction of the binary semi-major axis (see Table~\ref{tab:hydro_comparison}).

The studies also differ in the strategy for exploring the parameter space and in the duration of the simulations. \citetalias{zrake_equilibrium_2021} and \citetalias{siwek_orbital_2023} run a discrete number of individual simulations. Each simulation concerns a particular binary eccentricity and mass ratio, that are kept fixed for the whole run. Each simulation lasts 2000 orbits and 10\,000 orbits, respectively. At the end of the run, the forces acting on the binary and other quantities of interest are computed and averaged over the last hundreds of orbits. \citetalias{zrake_equilibrium_2021} exclusively explore the $q=1$ regime, but do so with 32 simulations and a fine eccentricity spacing. In comparison, \citetalias{siwek_orbital_2023} conduct nine simulations at $q=1$.

\citetalias{dorazio_orbital_2021} took a distinct approach, running a single extended simulation lasting 20\,000 orbits. In this simulation, eccentricity is gradually increased from $e=0$ to $e=0.9$. In this work we only consider the part with $e \leq 0.8$ in order to be consistent with the samples taken from the other two studies, and because for $e > 0.8$ the size of the mini-disks at the periapsis passage becomes smaller than the gravitational smoothing length, as noted by the authors.
The resulting $a'$ and $e'$ are then smoothed over an $\approx 450$-orbits window 
to average over the short-term variability and produce a result equivalent to having run $\approx 40$ simulations with eccentricity spaced by $\Delta e=0.02$.
As a consistency check, \citetalias{dorazio_orbital_2021} have run additional simulations at a fixed orbital configuration, and compared them with the values obtained through the eccentricity sweep, finding  a negligible difference in most cases. The advantage of the approach by \citetalias{dorazio_orbital_2021} is that, since the eccentricity is increased slowly, the disk is kept close to a relaxed state at all times, preventing the need to spend computational resources in relaxing a new disk for each eccentricity.

\subsection{Interpolation of the derivatives} \label{sec:interpolation}
To compute the long-term orbital evolution of the binary systems, we linearly interpolate the derivatives $a'$, $e'$ and $q'$ (as defined in Equation~\ref{eq:def_derivatives}) obtained from the hydrodynamic models. 
To this aim, we define a set of interpolating functions $f_a(e,q)$, $f_e(e,q)$, and $f_q(e,q)$. These functions do not depend on the separation $a$ because of the scale invariance of the problem.
For our fiducial model, based on \citetalias{siwek_orbital_2023}, we use bilinear interpolation\footnote{\url{https://docs.scipy.org/doc/scipy/reference/generated/scipy.interpolate.RegularGridInterpolator.html}} \citep{weiser_note_1988}
on the rectangular grid of eccentricities and mass ratios to obtain the functions $f_a(e,q)$ and $f_e(e,q)$, which interpolate the values provided in Tables~\ref{tab:adota_siwek} and \ref{tab:edot_siwek}. These are the same values reported in Figs.~2 and 3 of \citetalias{siwek_orbital_2023}, with an additional suite of simulations at $e=0.05$, previously unpublished.

We obtain $f_q(e,q)$ first by interpolating the relative accretion rates $\lambda\equiv\dot M_2 / \dot M_1$  published in Fig.~4 of \citetalias{siwek_preferential_2023} (also provided in our appendix in Table~\ref{tab:lambda_siwek} for convenience), then by computing
\begin{equation}
    f_q(e,q) \equiv \frac{(1+q)(\lambda(e,q) - q)}{1+\lambda(e,q)}.
\end{equation}
Our functions $f_a$, $f_e$ and $f_q$ are defined for $0 \leq e \leq 0.8$ and $0.1 \leq q \leq 1$. In case the integration routine requires to evaluate them outside these ranges, we return the value of the nearest known data point.

We tested a cubic spline and the \texttt{PCHIP} method \citep{fritsch_method_1984} as alternative interpolation techniques. We observed that using a cubic spline introduced spurious oscillations in the interpolating functions. Instead, the \texttt{PCHIP} method, which is guaranteed to preserve the monotonicity of the data, gave satisfactory results that closely resembled linear interpolation. Since \texttt{PCHIP} and the linear method gave equivalent results, we opted to use the simpler linear interpolation.

For \citetalias{zrake_equilibrium_2021} and \citetalias{dorazio_orbital_2021} we only have to consider one dimension: eccentricity.  In the case of \citetalias{zrake_equilibrium_2021}, the derivative of the semi-major axis is not provided in their work. We publish it here, together with the eccentricity derivative, and provide them in Table~\ref{tab:derivatives_Z21_DD21}, in Appendix~\ref{app:tables_of_derivatives}.
In the case of \citetalias{dorazio_orbital_2021}, their eccentricity sweep yielded continuous curves for the derivatives $a'$ and $e'$ instead of discrete points. Their results show oscillations (see Fig.~1 in \citetalias{dorazio_orbital_2021}), even after their smoothing procedure. As we expect that these oscillations do not represent the steady state, we remove them by discrete resampling with carefully selected points to preserve the major features as shown in Fig.~\ref{fig:DD_derivatives}. The results are reported in Table~\ref{tab:derivatives_Z21_DD21}.
We then obtain our interpolating functions for \citetalias{zrake_equilibrium_2021} and \citetalias{dorazio_orbital_2021} by linear interpolation of these tables. 

\subsection{Integration of the orbital evolution} \label{sec:integration}
The evolution of the binary system is then computed by solving the differential equations

\begin{align}
\label{eq:a_evolution}
a' \equiv \diff{\log a}{\log M} &= f_a(e,q), \\ 
\label{eq:e_evolution}
e' \equiv \diff{e}{\log M} &= f_e(e,q), \\
\label{eq:q_evolution}
q' \equiv \diff{q}{\log M} &= f_q(e,q),
\end{align}

for the variables $a$, $e$, and $q$, with appropriate initial conditions $e_0$ and $q_0$. Given the scale invariance of the problem, we always set $a_0=1$. The integration was performed with the \texttt{LSODA}  method \citep{petzold_automatic_1983} as implemented in SciPy. The relative tolerance (\texttt{rtol}) was set at $10^{-2}$, and the absolute tolerance (\texttt{atol}) at $10^{-5}$.

\section{Results} \label{sec:results}

\subsection{Individual binary evolution}
\label{sec:trajectories}

\begin{figure*}
    \centering
    \includegraphics[width=\textwidth]{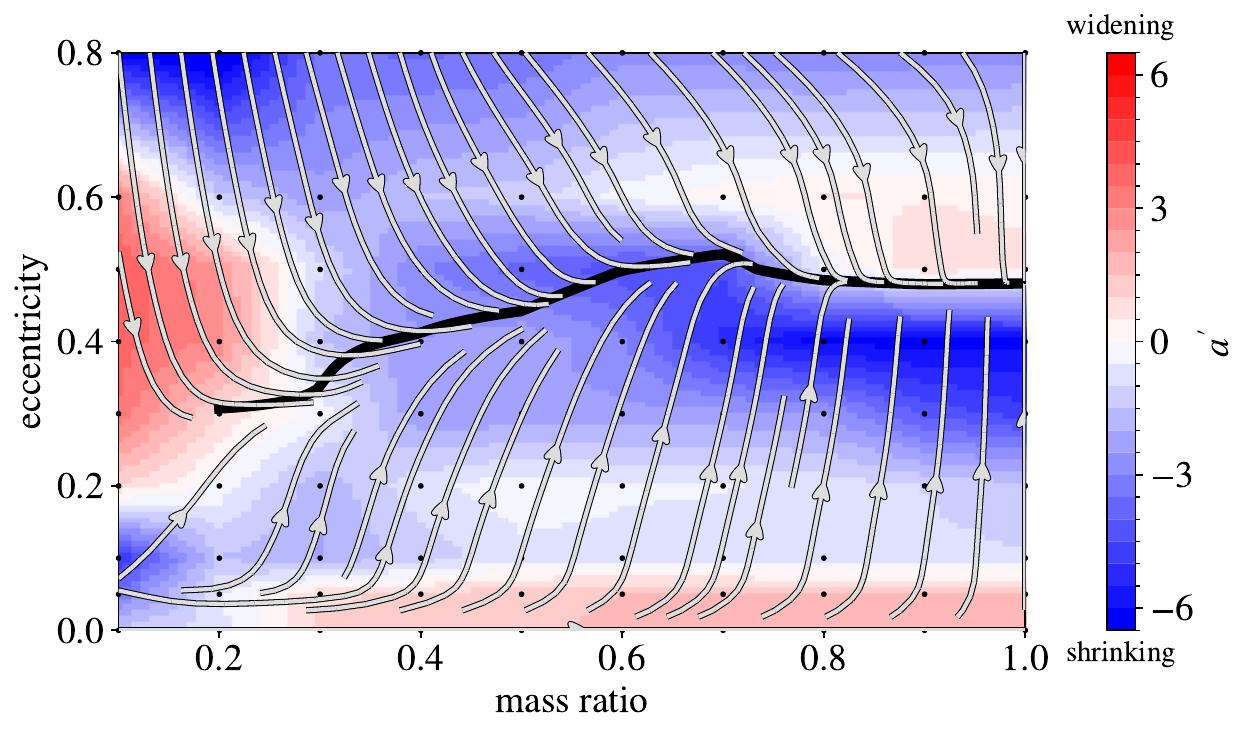}
    \caption{Orbital evolution of binary systems under interaction with a circumbinary disk. The white lines follow the trajectories in the eccentricity--mass ratio plane. The thick black line traces the equilibrium eccentricity, that is the region where, for a fixed mass ratio $q$, the eccentricity derivative is $e'=0$. The background color shows our interpolated function approximating the derivative of the separation $a'\equiv d\log a/d\log M$.  This quantity shows how fast the separation changes per unit of accreted mass. Regions where the orbit shrinks are shown in blue and regions where the orbit widens in red. The small black dots in the background correspond to the parameters used in the hydrodynamic simulations by \citetalias{siwek_orbital_2023}, on which the interpolation is based.
    }
    \label{fig:streamplot}
\end{figure*}

The evolution of a binary interacting with and accreting from a circumbinary disk is captured by the derivatives $a'$, $q'$ and $e'$ (Equations \ref{eq:a_evolution}, \ref{eq:e_evolution} \& \ref{eq:q_evolution}), which describe the change of each of these parameters as a function of the accreted mass. 

In Fig.~\ref{fig:streamplot}, we show our interpolation for the evolution of the semi-major axis as a function of  eccentricity and mass ratio.
The red regions denote values of $q$ and $e$ where the disk increases the binary separation, while the blue regions mark where it causes the binary separation to shrink.
The white lines with arrows show the trajectories traced by example systems as they evolve in $q$-$e$ space. 
The black band in the plot center denotes the equilibrium eccentricity as a function of $q$ (see \citetalias{siwek_orbital_2023}). 
All paths first move toward the equilibrium eccentricity, after which they tend to grow their mass ratio toward unity.
In the limit of infinite mass accretion, all streamlines eventually converge on a single attractor, with a mass ratio $q_{\rm att} = 1$ and eccentricity $e_{\rm att} \approx 0.5$.
At the attractor, binaries tend to shrink their orbit at a rate $a'_{\rm att} = -0.7$, or equivalently, like $a \propto M^{-0.7}$. 

\begin{figure}
    \centering\includegraphics[width=\columnwidth]{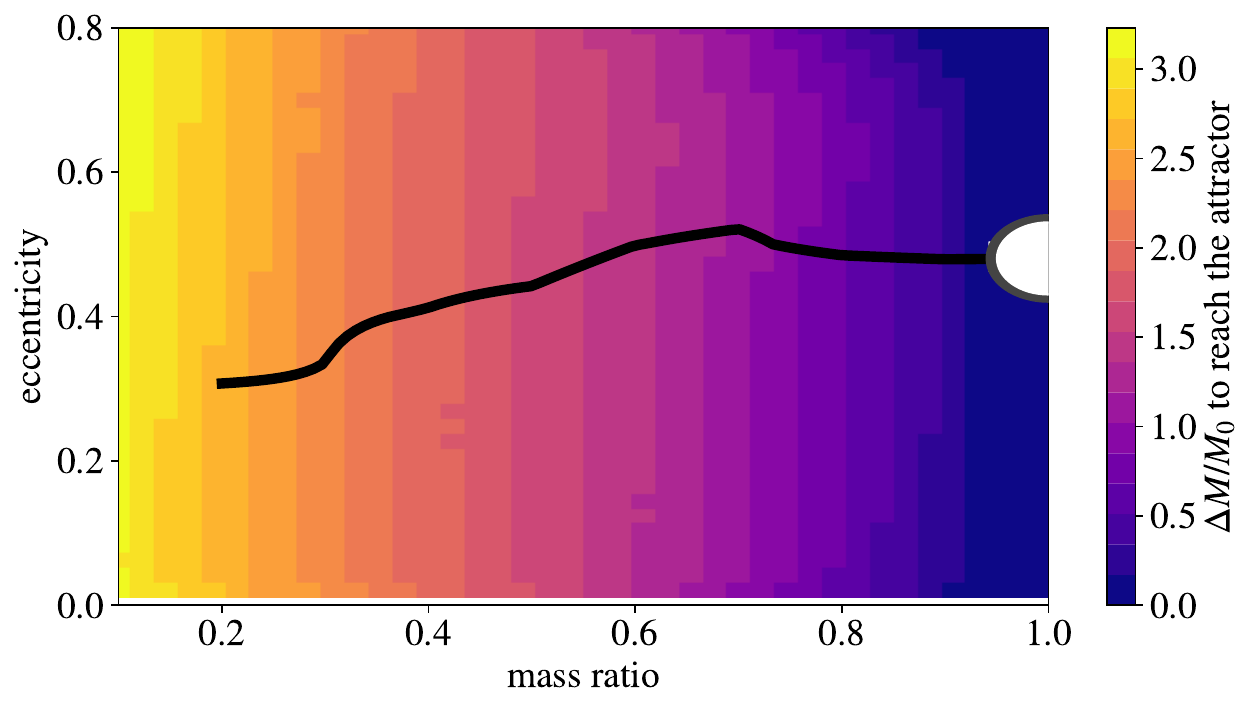}
    \centering\includegraphics[width=\columnwidth]{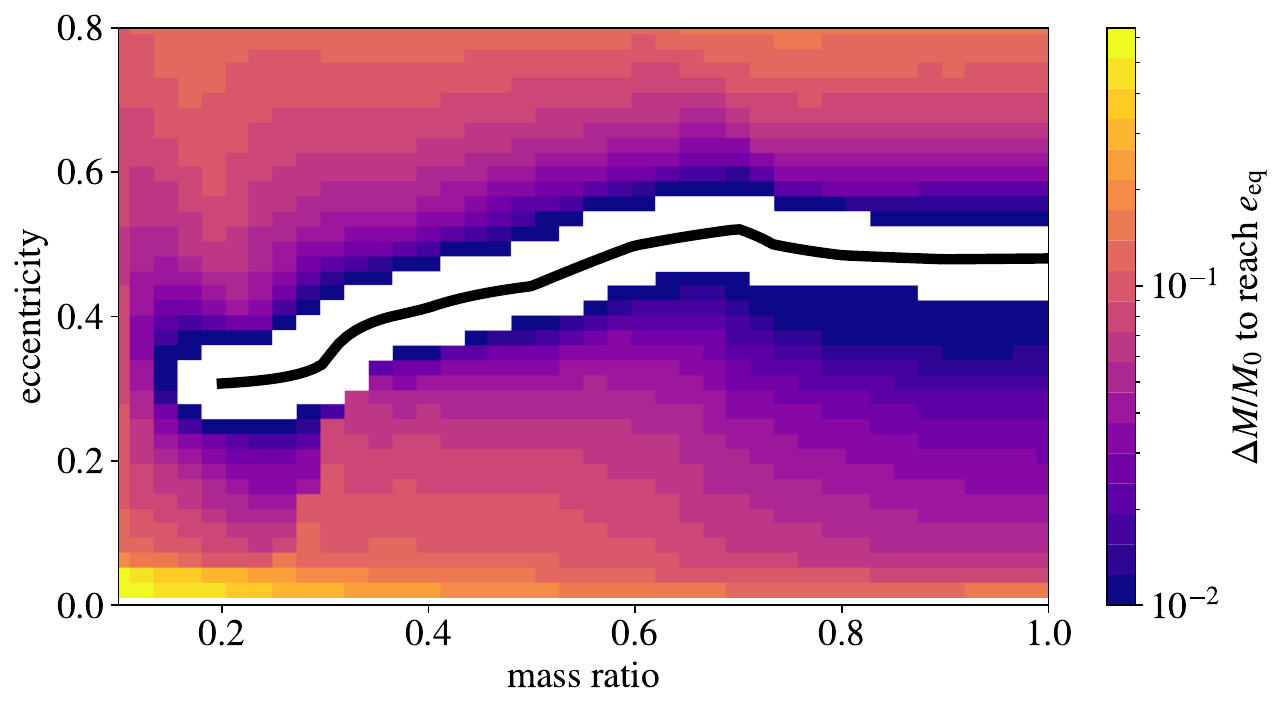}
    \caption{Effect of accretion on the orbit of the binary. The top panel shows how much mass the systems need to accrete to reach the vicinity of the attractor point (white circle). The bottom panel shows the amount of accreted mass needed to reach the equilibrium band only (white region). 
    The black line traces the equilibrium eccentricity. See also Figure \ref{fig:streamplot}.
    }
    \label{fig:mass_to_attractor}
\end{figure}

In a realistic scenario, systems may not reach the attractor because the mass reservoir feeding the disk is depleted or because other mechanisms begin to dominate the evolution of the orbit, such as stellar tides or gravitational-wave emission.
In Fig.~\ref{fig:mass_to_attractor} we quantify how much mass a system needs to accrete in order to approach both the equilibrium eccentricity and the final attractor, in the bottom and top panels, respectively.
Systems typically reach the equilibrium band after accreting less than $10\%$ of their initial mass, while they may require up to three times the initial mass to reach the attractor point at $q=1$.

\begin{figure}
    \centering\includegraphics[width=\columnwidth]{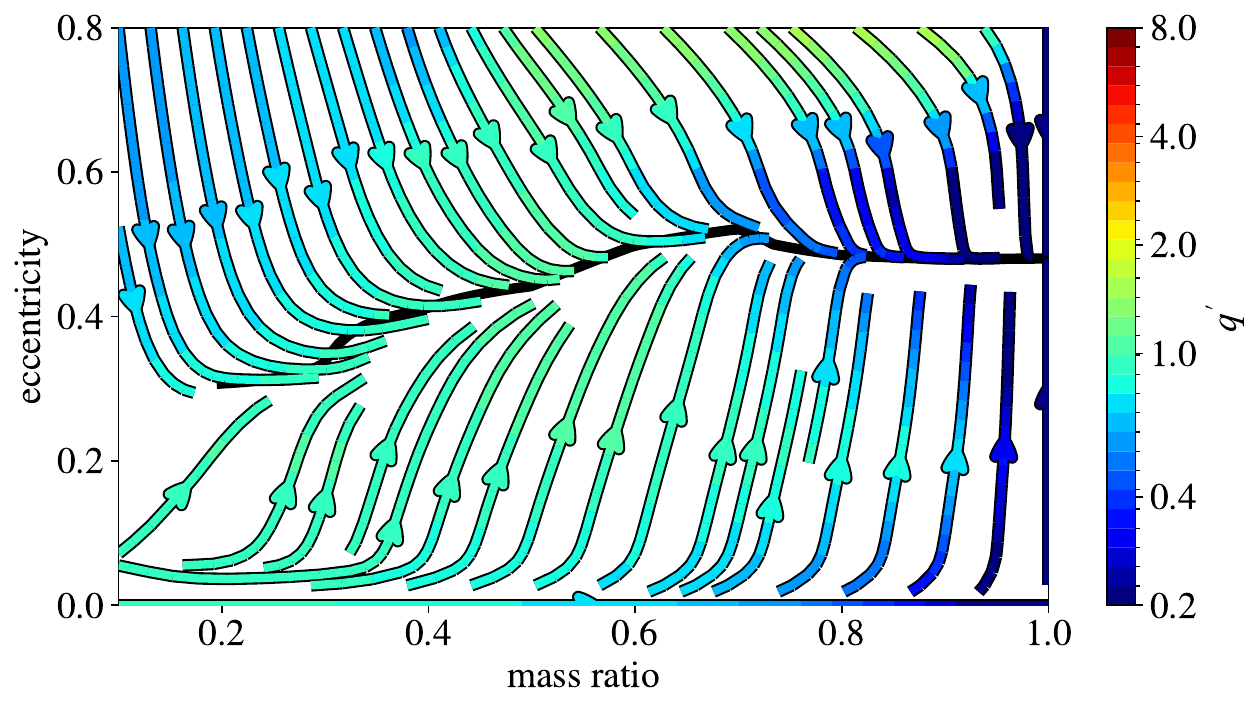}
    \centering\includegraphics[width=\columnwidth]{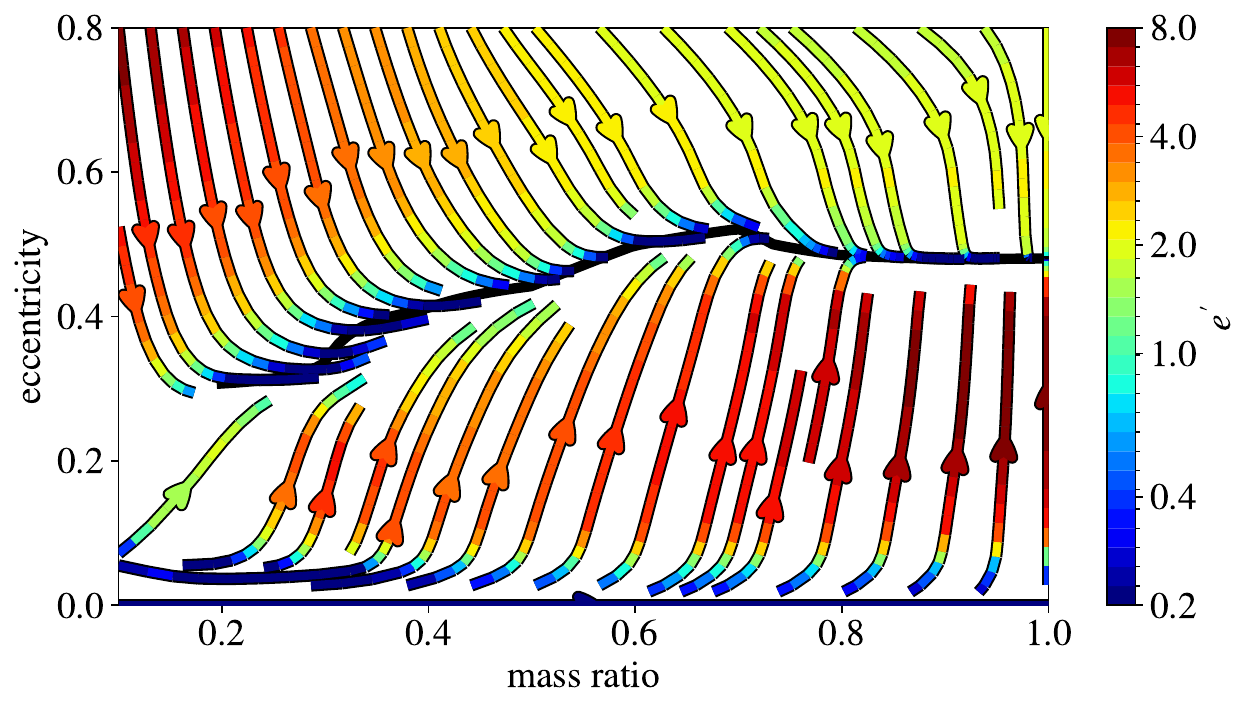}
    \caption{Velocity at which systems travel across the eccentricity--mass ratio plane. The streamlines are the same as in Figure \ref{fig:streamplot}, but here the color of the streamlines encodes how fast the mass ratio (top) and eccentricity (bottom) change per unit of accreted mass, quantified by the derivatives $|q'|$ and $|e'|$.
    The black line traces the equilibrium eccentricity.
    }
    \label{fig:edot_qdot}
\end{figure}

Binaries tend to adjust their eccentricity toward the equilibrium value before meaningfully altering their mass, as shown by the nearly vertical arrows in Fig.~\ref{fig:streamplot}.
This is because the magnitude of $e'$ is several times larger than $q'$.
This is shown in more detail for the same streamlines in Fig.~\ref{fig:edot_qdot} where the coloring shows the evolution rates for $q$ and $e$ in the top and bottom panels respectively.
It is important to note the difference in the dynamic range ($\max |e'| \sim 9$ while $\max |q'| \sim 1.5$), which can be read from the color bar. 
Figure~\ref{fig:edot_qdot} shows that the systems whose eccentricity evolves faster are those with nearly equal masses and low eccentricities ($0.8 < q < 1$, $0.2 < e < 0.4$), and those with large eccentricities and mass ratios ($q < 0.2$, $e > 0.6$). The least sensitive systems are nearly circular ($e=0-0.1$) and those near the equilibrium eccentricity band. 
There also exists an asymmetry for the behavior of systems with near equal masses ($q>0.8$) in that the pumping of $e$ at initially modest eccentricities ($0.1 < e < 0.4$) is much stronger than the damping in systems with initially high $e \gtrsim 0.5$ (see also the discussions in \citealt{dunhill_2015} and \citetalias{siwek_preferential_2023}).

Generally, only when binaries have functionally reached the equilibrium eccentricity (or are almost perfectly circular) does the evolution of $q$ start to dominate, $|q'| > |e'|$.
As a result, the evolution of a typical system is composed of two phases. In the first phase, eccentricity evolves toward the equilibrium band, keeping the mass ratio almost constant. In the second phase, the mass ratio moves toward unity, while the eccentricity tracks the $q$-dependent equilibrium value.

\begin{figure}
    \centering
    \includegraphics[width=\columnwidth]{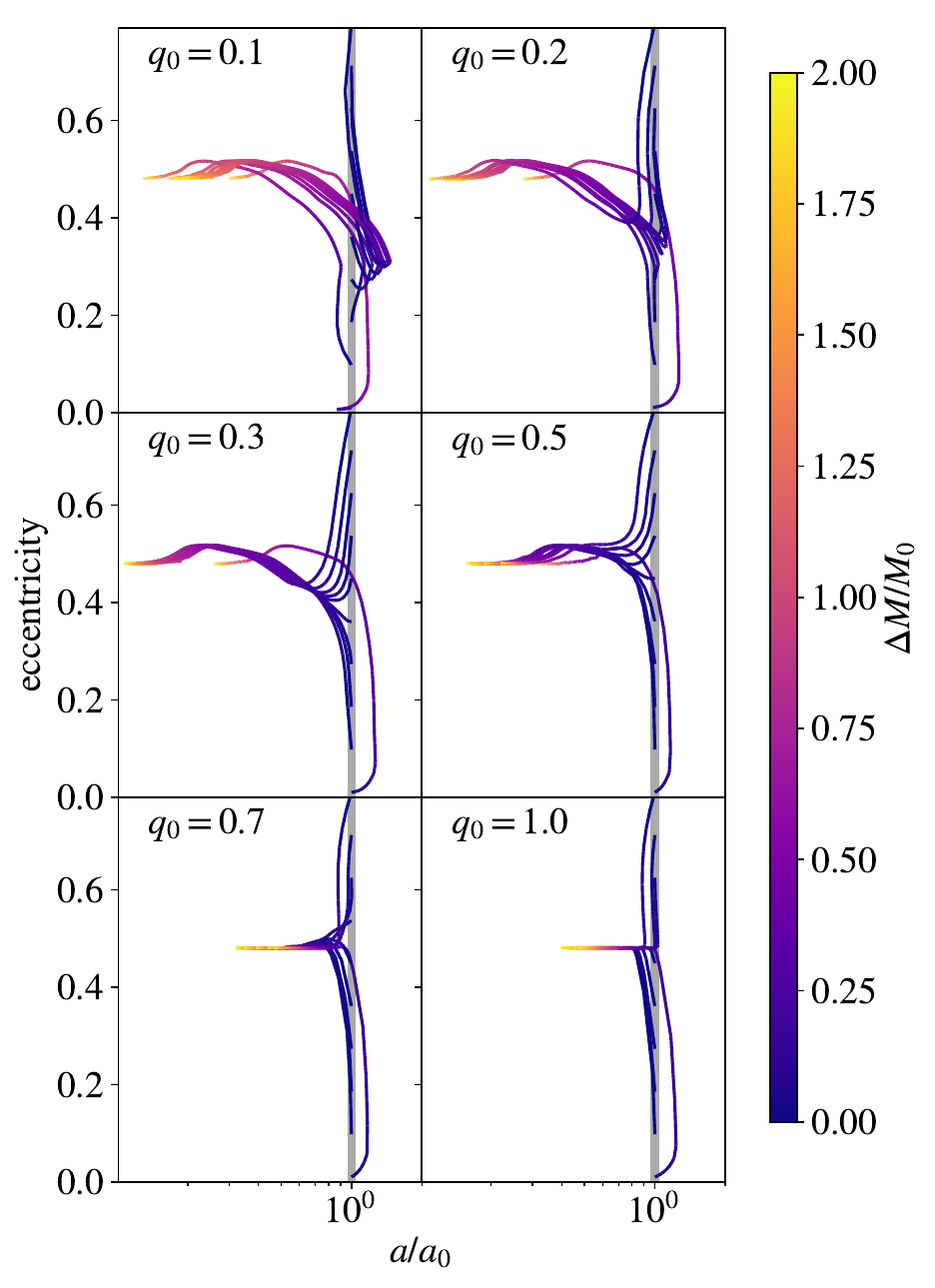}
    \caption{Evolution of eccentricity ($e$) and semi-major axis ($a$) as a function of accreted mass. For each panel, the systems start from a different initial mass ratio ($q_0$) that spans from $0.1$ (top left) to $1.0$ (bottom right). Within each panel, the initial eccentricity ($e_0$) spans from $0.01$ to $0.8$. The trajectories are colored based on the mass change $\Delta M/M_0$, where $M_0$ is the initial mass of the binary and $\Delta M$ is the amount of accreted mass. All the systems start along the vertical line $a=a_0$, highlighted with a thick gray line.
    }
    \label{fig:evolve2_siwek}
\end{figure}

Figure~\ref{fig:evolve2_siwek} shows the evolution of the eccentricity and semi-major axis of a series of accreting binary systems. Each system is initialized with a different eccentricity along the vertical gray bar on the right side of each panel, and the initial mass ratio is noted in each panel's upper left corner. 
As mass is accreted on to the binary, the eccentricity again changes most rapidly such that each system initially reaches the equilibrium value after accreting $\approx 10\%$ of its original mass, and then subsequently only varies with the q-dependence of the equilibrium eccentricity as the binary mass ratio grows.

Despite the presence of several regions of the $e$-$q$ space with a positive $a'$, in the region of equilibrium eccentricity
$a'$ is almost always negative, and the orbit will tend to shrink. However, similar to the mass ratio case, in order to achieve a significant reduction of the orbital separation $|\Delta a| \approx a_0$, the system needs to accrete from the disk an amount of mass comparable to the initial mass of the system $\Delta M \approx M_0$. 
This implies that orbital decay from a steady, disk-mediated accretion phase may only be viable in the limit that the gas reservoir is at least of comparable mass to the binary itself.

\subsection{Population evolution} \label{sec:population}

In addition to examining the evolutionary trajectories of individual accreting systems, we also explore the properties of populations and describe the observational signatures that they produce on the distributions of orbital parameters. 
We randomly sample a population of binary systems from a flat distribution of eccentricity between $0 \leq e_0 \leq 0.8$ and a flat distribution of mass ratio between $0.1 \leq q_0 \leq 1$. Because the problem is scale-free, 
all systems are initialized with a semi-major axis of $a_0 = 1$.
Then, assuming each system accretes at the same rate, we model our population forward in time according to the solutions of Equations~\ref{eq:a_evolution}-\ref{eq:q_evolution}.
We choose these simple initial conditions to be agnostic to the type of binary systems (stars, black holes, etc.) that we are modeling. 
We sample a population of $10^6$ systems, which we deem large enough to densely cover the two-dimensional parameter space of eccentricity and mass ratio.

\begin{figure}
    \centering
    \includegraphics[width=\columnwidth]{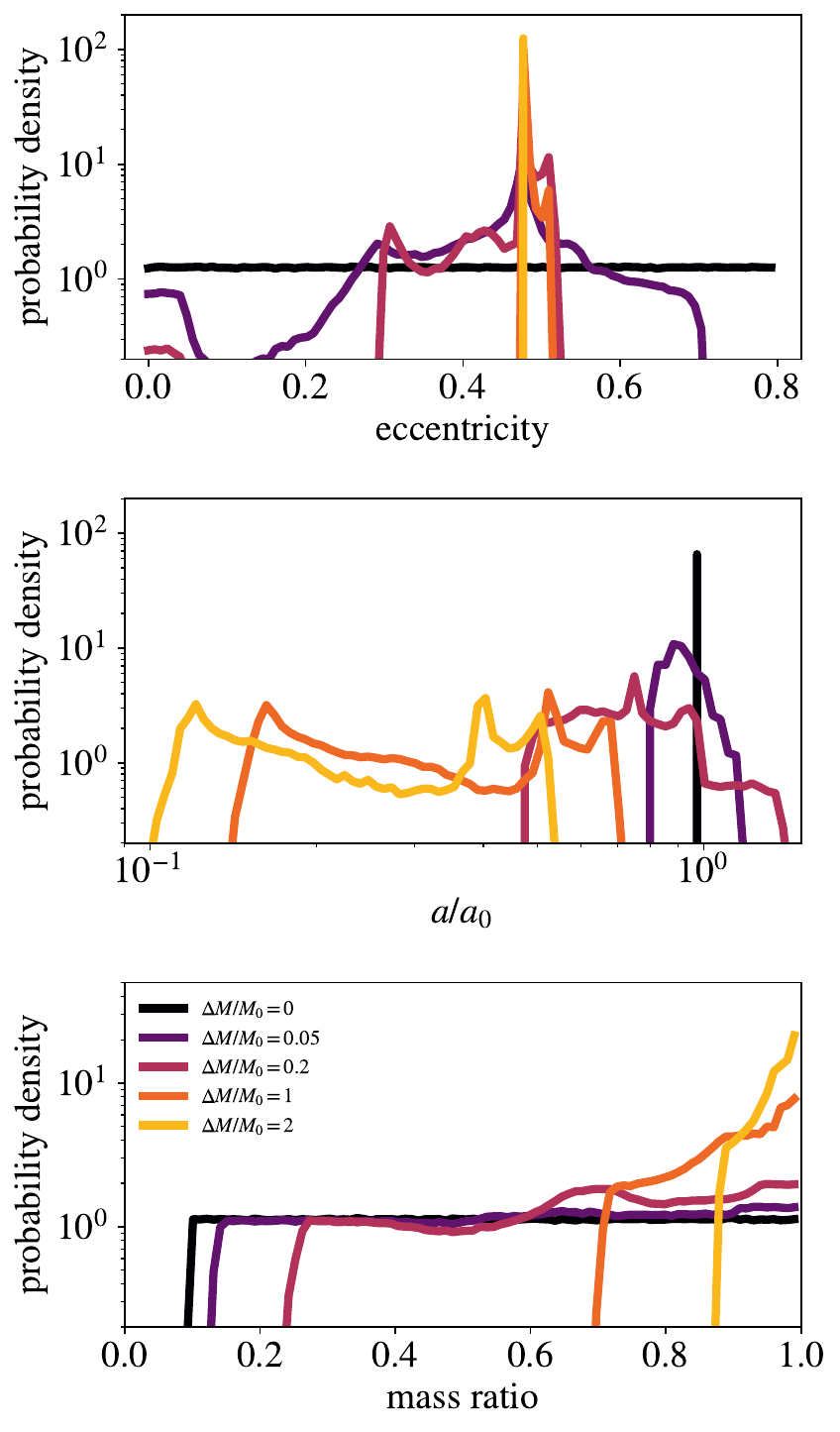}
    \caption{Evolution of the distribution of the orbital parameters for a population of binaries. Distribution of eccentricity (top), semi-major axis (middle), and mass ratio (bottom). Each color corresponds to a different amount of accreted mass, increasing from dark purple to light yellow.}
    \label{fig:population_siwek}
\end{figure}

Figure~\ref{fig:population_siwek} shows the population distributions of eccentricity, semi-major axis, and mass ratio at different accreted masses $\Delta M$.
Up to the accretion of $\approx 10\%$ of the initial mass ($\Delta M/M_0 \leq 0.1$), the most noticeable variation happens in the eccentricity distribution in the top panel, which concentrates around $e \approx 0.5$
After accreting more than $50\%$ of the initial mass, however, the mass ratio and semi-major axis distributions begin to change noticeably. 
The mass ratio distribution develops a peak at $q=1$ as the initially largest-$q$ systems grow to equal mass, and mass ratios below $q=0.5$ are depopulated. The distribution of $a/a_0$ broadens to $\approx 0.6$ dex and features a double peak. The rightmost peak is composed of those systems that start with $q>0.8$, which experience a slow evolution of the semi-major axis due to a small value of $a'(e{=}e_{\rm eq}) \approx -0.7$. On the contrary, the systems that start with $q<0.8$ can have $a'(e{=}e_{\rm eq})$ as large as $-4$ (see also Fig.~\ref{fig:eq_ecc_adot} in Appendix~\ref{app:additional_figures}) so they migrate to a tighter orbit faster, and eventually accumulate on the leftmost peak upon reaching equal mass ratio.

\begin{figure}
    \centering
    \includegraphics[width=\columnwidth]{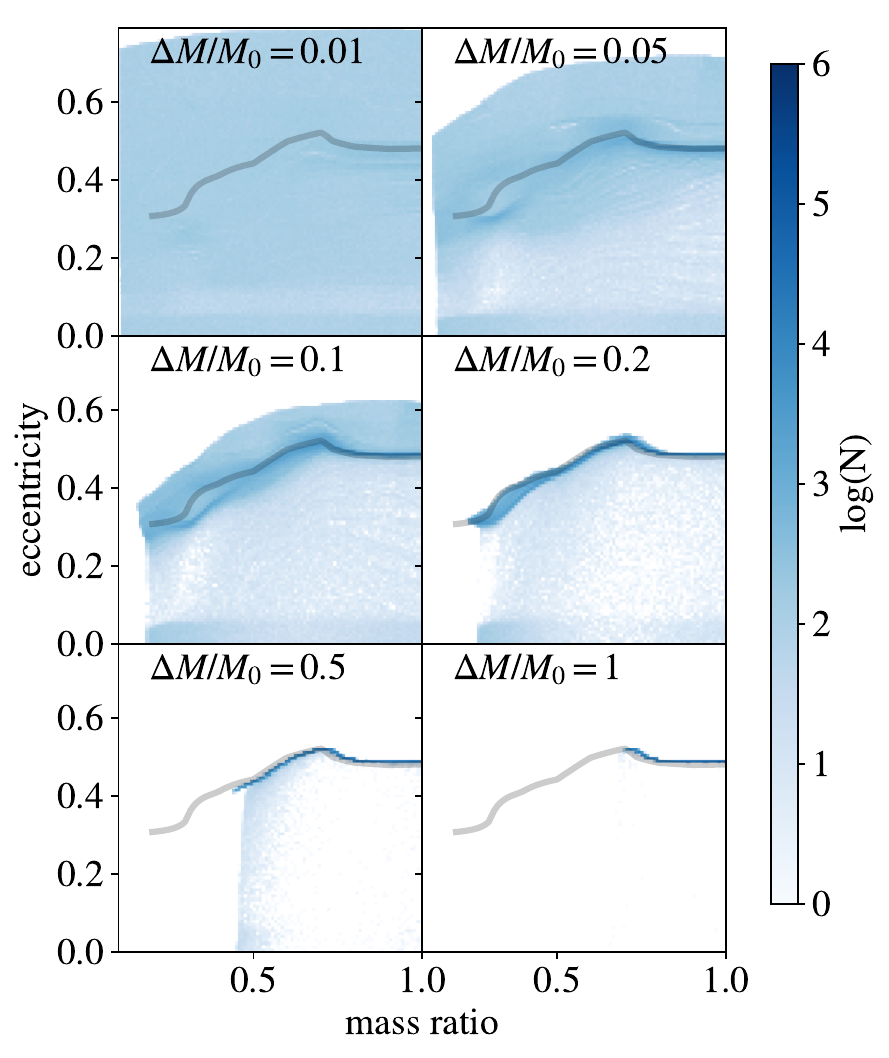}
    \caption{Population synthesis of binaries interacting with circumbinary disks in the $e$-$q$ plane. Each panel corresponds to a different amount of accreted mass, increasing from top-left to bottom-right. The color scale indicates the number of systems falling into each bin. The gray band marks the position of the equilibrium eccentricity.
    }
    \label{fig:e_q_correlation}
\end{figure}

Figure~\ref{fig:e_q_correlation} depicts the evolution of the population in the $e$-$q$ plane. In the top-left panel, every system has accreted 1\% of the initial mass, and they remain close to their initially flat distributions because 1\% is not sufficient to alter the distribution of eccentricity or mass ratio.
After accreting 5\% of the initial mass --- in the top-right panel --- a pattern starts to emerge and becomes evident in the middle-left panel as a clear correlation between eccentricity and mass ratio, caused by the accumulation of systems close to their equilibrium eccentricity. When the systems have accreted 20\% of the initial mass (middle-right panel) the $e$-$q$ correlation becomes extremely tight. In the remaining two panels, after accreting 50\% or 100\% of the initial mass, all the systems settle at the same eccentricity $e\approx 0.5$ and move toward a mass ratio of one. This $e$-$q$ correlation could be observed in populations of binaries that have undergone steady, disk-mediated accretion in the past, as long as additional dynamics have not since meaningfully modified the orbit. The presence of the correlation could also help constrain the typical amount of mass that was accreted from the disk.

Because of the long real-time evolutionary timescales of the relevant stellar or black hole binaries, identifying (or ruling out) features in parameter distributions of populations is likely the most promising approach for determining and constraining the importance of accretion phases in binary evolution.

\section{Comparison with other models} \label{sec:comparing_results}

Here we compare the results of our fiducial model to those based on the additional simulations discussed in Sect.~\ref{sec:hydro_comparison}.
The predictions of the three models broadly agree, all of them developing an equilibrium eccentricity around $e \approx 0.4 - 0.5$. 
An intuitive explanation for the existence of an equilibrium eccentricity can be understood in terms of the interaction between the inner edge of the disk and the less massive component of the binary, as described by \citet{artymowicz_1991} and \citet{roedig_limiting_2011}. Namely, when the binary is at apoapsis, the less massive object approaches the inner disk edge. If its angular velocity surpasses that of the fluid in the disk, it is decelerated, leading to an increase in orbital eccentricity. Conversely, when the object's angular velocity lags behind that of the disk fluid, the object is propelled forward, subsequently reducing the orbital eccentricity. Equilibrium is approximately achieved when the angular velocities of the object at apoapsis and the disk fluid are the same.
We note, however, that this intuitive picture is approximate and does not fully account for the disk's morphological complexities in the full nonlinear solutions. For instance, the equilibrium eccentricity in \citetalias{dorazio_orbital_2021} does not conform to this picture, but is associated with the disk transitioning from a symmetric to an asymmetric state.

We report here (and in Table~\ref{tab:hydro_comparison}) the equilibrium eccentricity for the three models at $q=1$, obtained via linear interpolation. We find $e_{\rm eq} = 0.48^{+0.02}_{-0.08}$ for \citetalias{siwek_orbital_2023}, $0.443^{+0.007}_{-0.018}$ for Z21, and $0.39 \pm 0.01$ for \citetalias{dorazio_orbital_2021}.
The uncertainty is taken to be the distance from the closest known point, except in the case of \citetalias{dorazio_orbital_2021}, where it is half of the reported eccentricity resolution $\Delta e=0.02$. These results are consistent with the ones reported in the respective papers.
See also Fig.~\ref{fig:derivatives_q1} for a comparison of the derivatives obtained by the three suites of simulations.

\subsection{Behavior at small eccentricity} \label{sec:behaviour_small_e}
The primary evolutionary difference between the three suites of simulations is the sign of $e'$ for $e \lesssim 0.1$. 
\citetalias{zrake_equilibrium_2021} and \citetalias{dorazio_orbital_2021} both determine that such near-circular orbits are stable and remain near-circular with $e' < 0$, while \citetalias{siwek_orbital_2023} find that such near-circular orbits are unstable with $e' > 0$ always.
This small discrepancy implies a dramatically different evolution for systems starting from nearly circular orbits. According to \citetalias{siwek_orbital_2023}, those systems will increase their eccentricity up to the equilibrium value $e \approx 0.5$ and then shrink the orbit, while for \citetalias{zrake_equilibrium_2021} and \citetalias{dorazio_orbital_2021}, systems with small eccentricity will tend to circularize while they grow their orbital separation.
A positive $e'$ for $e \le 0.2$ was also found by SPH studies \citep[e.g.,][]{artymowicz_1991, cuadra_2009, roedig_limiting_2011, fleming_coevolution_2017}. 

The reason for this discrepancy is currently unclear.
\citetalias{siwek_orbital_2023} attributes this discrepancy to different radial dependencies in the kinematic viscosity adopted, for instance by \citetalias{dorazio_orbital_2021}, but this cannot explain how \citetalias{siwek_orbital_2023} and \citetalias{zrake_equilibrium_2021}---who adopt the same viscosity prescription --- obtain a different result. 
Aside from the grid structure, the other major difference between these studies is that the mass removal rate in the sink-region (see Sect.~\ref{sec:hydro_comparison}) is much more rapid in \citetalias{siwek_orbital_2023} than in the two additional studies. However, some of us have examined whether this could account for the discrepancy and found that commensurately increasing the sink-rate in Eulerian grid-based simulations like \citetalias{dorazio_orbital_2021} and \citetalias{zrake_equilibrium_2021} actually exacerbates the disparity.
See Sect.~\ref{sec:application_post_ce} for a discussion of the consequences for post--common-envelope systems.

\subsection{Orbital decay}
\label{sec:orbital_decay}

\begin{figure*}
    \centering
    \includegraphics[width=\textwidth]{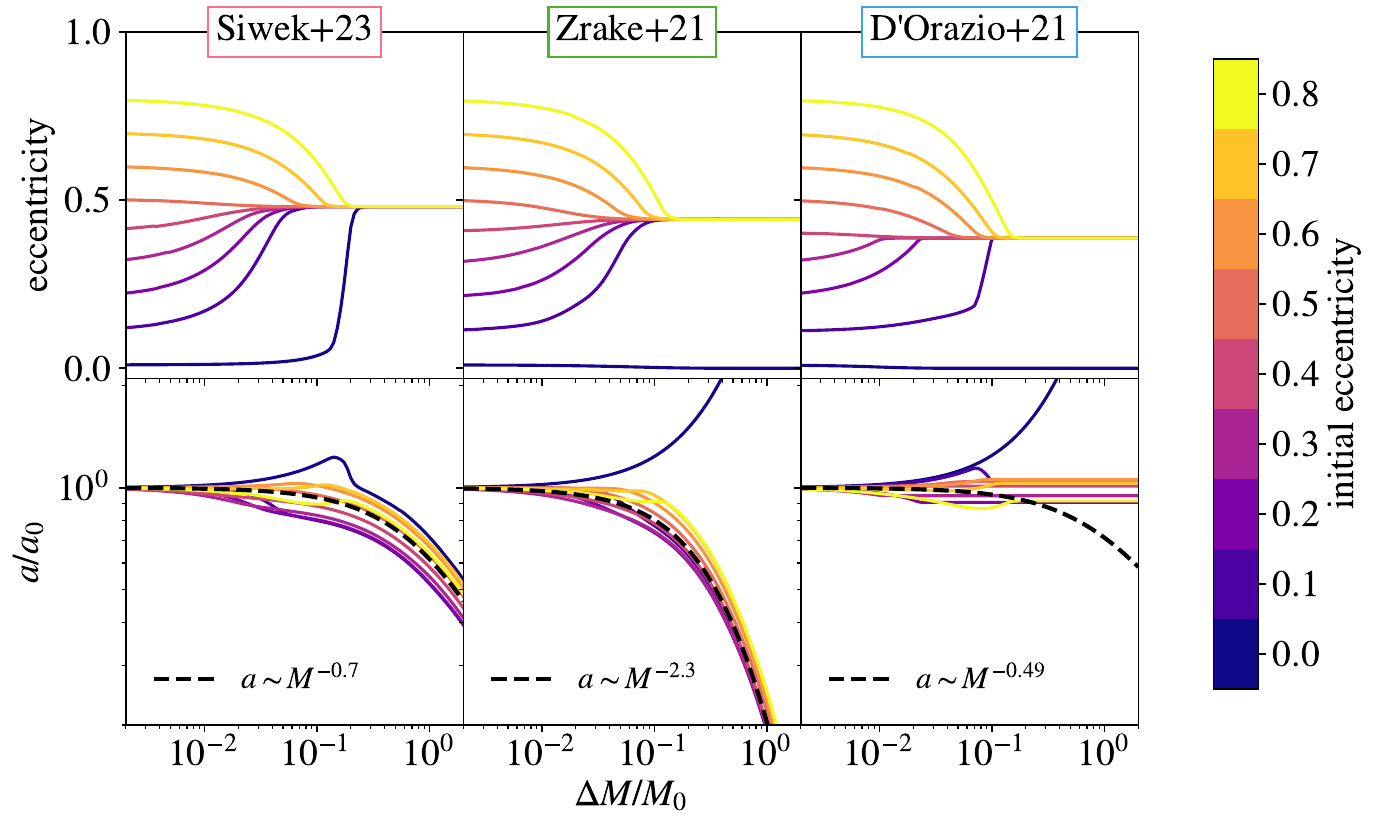}
    \caption{Model comparison of the orbital evolution of equal-mass binaries. We show eccentricity (top) and semi-major axis (bottom) as a function of accreted mass. Each line represents a different system, colored according to the initial eccentricity, in the range $0.01\leq e_0 \leq0.8$. The black dashed lines in the bottom panels mark the analytical solutions obtained in Sect.~\ref{sec:orbital_decay} considering a system evolving at the equilibrium eccentricity.}
    \label{fig:evolve2_comparison}
\end{figure*}

Figure~\ref{fig:evolve2_comparison} compares the long-term evolution of individual systems according to the three models, in the regime of $q=1$. The top row illustrates the change in eccentricity. Each system starts from a different initial eccentricity $e_0$, spanning from 0.01 to 0.8,  encoded by color. The three models agree that after accreting $\approx 10\%$ of the initial mass ($\Delta M/M_0\approx 0.1$), all systems converge toward an equilibrium eccentricity. The only exception is the system initialized with $e_0=0.01$ (dark blue line), which tends to circularize for \citetalias{zrake_equilibrium_2021} and \citetalias{dorazio_orbital_2021}, since $e'<0$ in those models.

The bottom row shows the evolution of the semi-major axis. The left panel confirms that, according to \citetalias{siwek_orbital_2023}, all the systems are led to orbital shrinking independently of the initial eccentricity. Conversely, the central and right panels show that for \citetalias{zrake_equilibrium_2021} and \citetalias{dorazio_orbital_2021}, the orbit shrinks only if the initial eccentricity is larger than about 0.1. The amount of accreted mass necessary to change the orbital separation is also model-dependent. This is due to the different values of $a'$ at the equilibrium eccentricity in the three models (see Table~\ref{tab:hydro_comparison}). The black dashed lines correspond to the case of a system evolving at the equilibrium eccentricity, whose evolution is given by the solution to the equation

\begin{equation*}
    \diff{\log a}{\log M} = \gamma\quad{\rm (constant)} \implies a \propto M^\gamma,
\end{equation*}
where $\gamma$ is the value of the derivative of the semi-major axis at the equilibrium eccentricity. The linear interpolation performed on the results of \citetalias{siwek_orbital_2023} and \citetalias{zrake_equilibrium_2021} gives $a'(e{=}e_{\rm eq})=-0.70$ and $a'(e{=}e_{\rm eq})=-2.3$, respectively. Conversely, \citetalias{dorazio_orbital_2021} report $a'(e{=}e_{\rm eq}) \approx 0$, which the authors explain with the presence of an abrupt transition in the disk geometry in correspondence with the equilibrium eccentricity. However, \citetalias{dorazio_orbital_2021} predict oscillations around the equilibrium configurations to yield an effective $a'(e{=}e_{\rm eq}) = -0.49$ (see also Fig.~\ref{fig:eq_ecc_adot}). We note that the present work does not take into account such oscillations when computing the orbital evolution, hence the mismatch between analytical and numerical prediction in the bottom right panel of Fig.~\ref{fig:evolve2_comparison}.

The different values of $a'(e{=}e_{\rm eq})$ obtained by the three models imply a different amount of orbital shrinking, for a given accreted mass. Namely, after doubling the mass of the system, we predict the semi-major axis to have shrunk by a factor $5$, $1.6$, and $1.4$, for \citetalias{zrake_equilibrium_2021}, \citetalias{siwek_orbital_2023} and \citetalias{dorazio_orbital_2021}, respectively.

\subsection{Population comparison} \label{sec:population_comparison}
\begin{figure*}
    \centering
    \includegraphics[width=\textwidth]{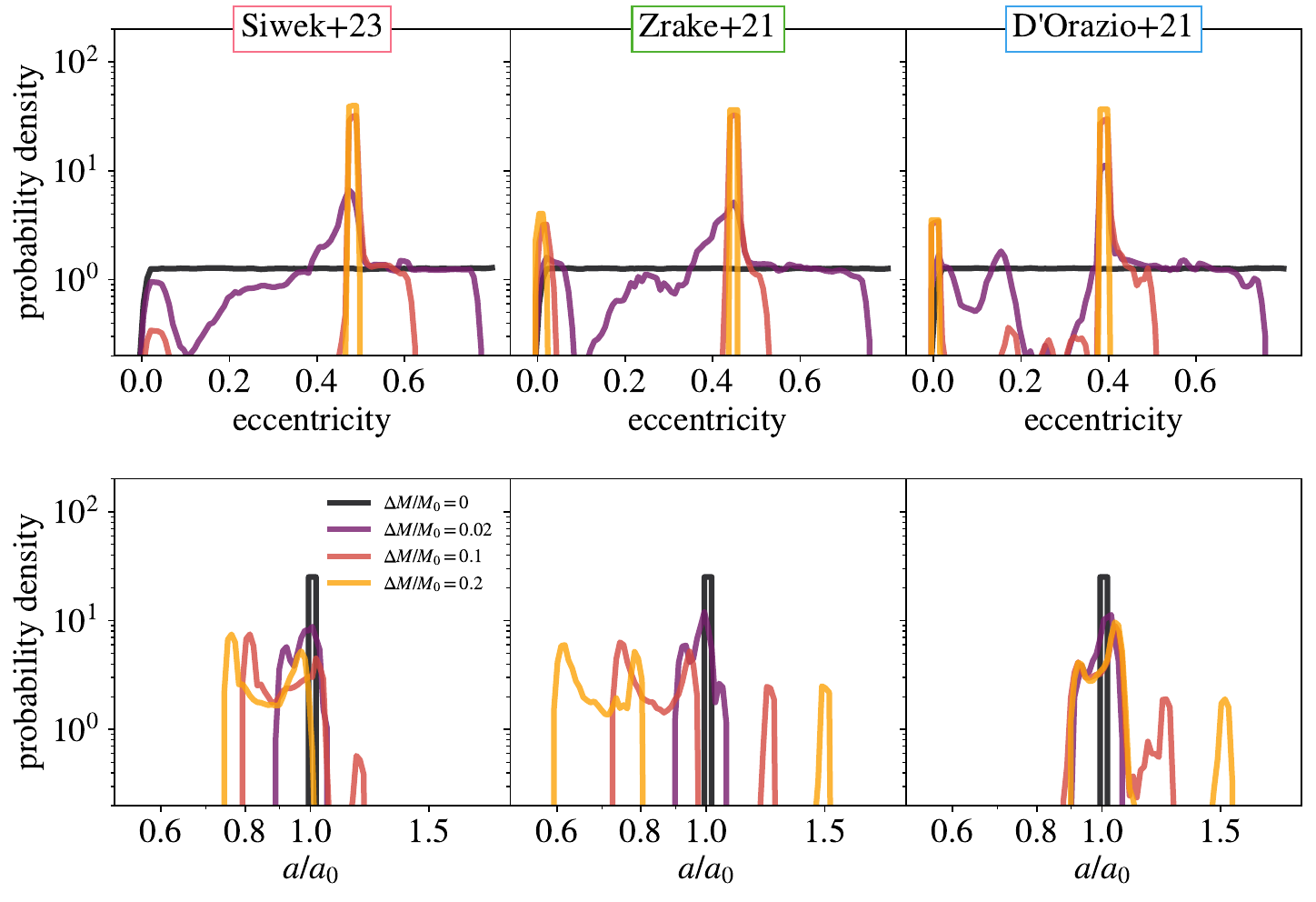}
    \caption{Model comparison of the evolution of the distributions of the orbital parameters for a population of equal-mass binaries. Distribution of eccentricity (top), semi-major axis (bottom), colored based on the amount of accreted mass. Each column shows one of the three models.
    }
    \label{fig:population_comparison}
\end{figure*}

Figure~\ref{fig:population_comparison} compares the distribution of the orbital parameters of a population of binary systems interacting with circumbinary disks, according to the three models. Here, similarly to Sect.~\ref{sec:population}, we have initialized $10^6$ systems with a flat distribution of eccentricity from $0$ to $0.8$ and a constant mass ratio $q=1$. The top row presents the distribution of eccentricity. For \citetalias{siwek_orbital_2023}, all the systems transition toward the equilibrium configuration, producing a visible peak in the eccentricity distribution already after the accretion of $2\%$ of the initial mass. The middle panel, relative to Z21, highlights that the eccentricity evolves toward a bimodal distribution peaked at $e=0$ and $e \approx 0.5$, with a gap at $e\approx 0.1$ (see also Figure~1 of \citetalias{zrake_equilibrium_2021}). The right panel shows the result for \citetalias{dorazio_orbital_2021}, which predicts an additional third peak, due to an accumulation of systems around $e\approx 0.2$. However, the third peak disappears when $\Delta M/M_0 \geq 0.1$.

The bottom panels show the distribution of the semi-major axis divided by the initial value $a_0$. All the systems start with $a/a_0=1$, but after having accreted some mass from the circumbinary disk, the semi-major axis distribution spreads. For \citetalias{siwek_orbital_2023}, all the systems are led to orbital shrinking, and this is reflected in the distribution shifting toward shorter semi-major axes. \citetalias{zrake_equilibrium_2021} presents a similar trend, except for the $\approx 10\%$ of the systems that started with $e_0<0.1$, so that their orbit is widening. The model based on \citetalias{dorazio_orbital_2021} also predicts the systems with small eccentricity to widen, while the others maintain approximately a constant semi-major axis. As discussed for Fig~\ref{fig:evolve2_comparison}, when considering small oscillations around the equilibrium eccentricity, the \citetalias{dorazio_orbital_2021} model would instead predict orbital shrinking.

\section{Astrophysical implications} \label{sec:implications}
We have investigated the long-term orbital evolution of binary systems subjected to the accretion of a circumbinary gas disk by employing the results of three suites of 2D hydrodynamic simulations. The scale-free nature of these simulations encompass a diverse range of astrophysical scenarios, including binary star formation (Section~\ref{sec:application_star_formation}), interacting binaries (Section~\ref{sec:application_interacting_stars}), post--common-envelope systems (Section~\ref{sec:application_post_ce}), post-AGB binaries (Section~\ref{sec:application_post_agb}) and SMBH (Section~\ref{sec:application_SMBH}). We discuss here the astrophysical and observational implications, and describe how these results can help guide the direction of future efforts and advancements in the field.

\begin{figure*}
    \centering
    \includegraphics[width=\textwidth]{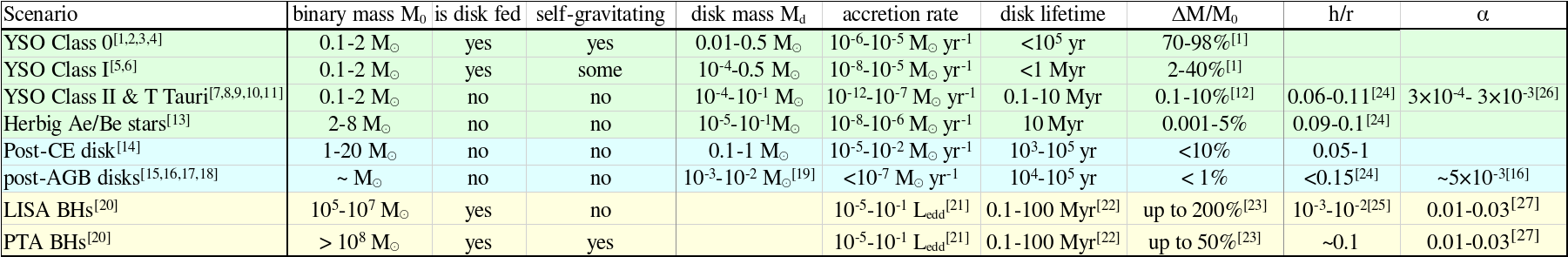}
    \caption{Summary of the properties of circumbinary disks in astrophysical scenarios. Binary mass refers to the initial mass $M_0$ of the central binary. The disk is fed if there is a mass inflow that replenishes the disk. The disk is self-gravitating if the self-gravity of the disk has a nonnegligible effect on the dynamics and structure of the disk. Disk mass $M_d$ is the instantaneous mass of the disk. Accretion rate is the mass accreted by the binary per unit time. Disk lifetime is the duration of a disk-mediated phase in the binary evolution, terminating when the disk is dispersed or the binary merges. Accreted mass is the amount of mass that can be accreted from the disk on to the binary within the disk lifetime. When the disk is not fed from external mass inflow, the accreted mass can be estimated as $M_d/M_0$, which assumes that all the mass in the disk is eventually accreted, and neglects outflows or other mechanisms of mass loss. $h/r$ is the disk aspect ratio in the vicinity of the binary. $\alpha$ is the viscosity parameter of the disk.
    \\
    \tableref{1}{\citealt{jorgensen_prosac_2009}},
    \tableref{2}{\citealt{tobin_vlaalma_2020}},
    \tableref{3}{\citealt{jacobsen_alma-pils_2018}},
    \tableref{4}{\citealt{kido_early_2023}},
    \tableref{5}{\citealt{fiorellino_mass_2023}},
    \tableref{6}{\citealt{takakuwa_angular_2014}},
    \tableref{7}{\citealt{manara_x-shooter_2020}},
    \tableref{8}{\citealt{guilloteau_gg_1999}},
    \tableref{9}{\citealt{scaife_long-wavelength_2013}},
    \tableref{10}{\citealt{dutrey_gg_2016}},
    \tableref{11}{\citealt{long_2021}},
    \tableref{12}{\citealt{testi_protoplanetary_2022}},
    \tableref{13}{\citealt{grant_dotm_2023}},
    \tableref{14}{\citealt{tuna_long-term_2023}},
    \tableref{15}{\citealt{vos_testing_2015}},
    \tableref{16}{\citealt{izzard_circumbinary_2023}},
    \tableref{17}{\citealt{bujarrabal_high-resolution_2018}},
    \tableref{18}{\citealt{kluska_population_2022}},
    \tableref{19}{\citealt{bujarrabal_extended_2013-1}},
    \tableref{20}{\citealt{franchini_circumbinary_2021}},
    \tableref{21}{\citealt{novak_feedback_2011}},
    \tableref{22}{\citealt{schawinski_active_2015}},
    \tableref{23}{\citealt{siwek_effect_2020}},
    \tableref{24}{estimated from Equation \ref{eq:aspect_ratio_star} assuming $r=\SI{100}{\au}$ and $\alpha=0.1$},
    \tableref{25}{estimated from Equation \ref{eq:aspect_ratio_smbh} assuming $r=\SI{0.01}{\parsec}$, $\alpha=0.1$ and $\eta=0.06$},
    \tableref{26}{\citealt{rosotti_empirical_2023}},
    \tableref{27}{\citealt{starling_constraints_2004}}.
    }
    
    \label{fig:astro_disks}
\end{figure*}

\subsection{Binary star formation} \label{sec:application_star_formation}
One proposed mechanism for the formation of close binary stars \citep[see][for a review]{offner_origin_2023} revolves around the concept of disk fragmentation \citep{adams_eccentric_1989, shu_sling_1990, bonnell_formation_1994, mignon-risse_disk_2023}. 
In this scenario, gravitationally unstable clumps in the proto-stellar disk undergo inward migration as they grow \citep{ward_survival_1997, kley_planet-disk_2012}, ultimately resulting in the formation of a central gas cavity around the two young stellar objects. 
After the formation of the cavity, if the orbital evolution is primarily governed by the interaction with a stable thin accretion disk, similar to the ones considered in this work, we expect to observe clear signs of this interaction in the eccentricity and mass ratio distributions among young main-sequence binaries. Particularly, we expect this signature to be evident in systems with sufficiently wide orbits, which have remained unaffected by tides.
According to our fiducial model, we expect a correlation between eccentricity and mass ratio, as illustrated in Figure~\ref{fig:e_q_correlation}. In contrast, the models based on \citetalias{zrake_equilibrium_2021} and \citetalias{dorazio_orbital_2021} predict a bimodal eccentricity distribution with peaks at $e=0$ and $e\approx0.4$, corresponding to the two equilibria.

Observationally, such Young Stellar Objects (YSOs) come with a variety of different disks---with varying adherence to the assumptions laid out in Sect.~\ref{sec:methods}---the most pertinent aspects of which we detail below and summarize in Table~\ref{fig:astro_disks}.
YSOs can be subdivided into classes \citep{dunham_evolution_2014} depending on the shape of their spectra, which also roughly correspond to progressive evolutionary states.

Class 0 YSOs are still too deeply embedded in gas to be detectable in the near infrared. Measuring their properties is challenging, but it is often possible to determine the presence of an accretion disk with a mass around 1-10\% of the envelope mass \citep{jorgensen_prosac_2009}.
In this regime, the disk is expected to be thick, self-gravitating, and possibly unstable to fragmentation \citep{kratter_gravitational_2016}. 
For example, the inward migration and subsequent accretion of fragments by the central star has been proposed as the mechanism driving the outbursts of FU Orionis stars \citep{hartmann_fu_1996, elbakyan_accretion_2021}.
This merger-dominated mass assembly has also been invoked as a likely pathway to the formation of binary stars \citep{tokovinin_formation_2020, riaz_turbulence_2021}.
Therefore, in the early phases of star formation, when accretion is bursty and dominated by mergers, the steady accretion models employed in this work may not be applicable.

In Class I YSOs (the next evolutionary stage), the envelope has largely dissipated and 70-98\% of the mass in the star-disk-envelope system is contained in the star itself \citep{jorgensen_prosac_2009}.
Therefore, assuming no mass inflows or outflows, we estimate a limit on the amount of mass that can still be accreted during star formation to $\Delta M/M_0 \approx 2-40\%$.
\citet{fiorellino_mass_2023} find that 20-30\% of the YSOs observed in this stage have a disk mass too small to trigger fragmentation, resulting in stable disks like the ones considered in this work. 
One of the few examples of a binary system observed in this configuration is L1551 \citep{takakuwa_angular_2014}.

Class II YSOs and T Tauri stars correspond to the next evolutionary stage.
At this stage, the envelope has completely dissipated and it is usually assumed that there is no significant inflow to feed the disk \citep[although recent findings are challenging this assumption, see][]{gupta_reflections_2023, kuffmeier_rejuvenating_2023}. 
As a consequence, the mass accreted in this stage is limited by the mass of the disk, which is typically $10^{-4}-10^{-1}\,\si{\Msun}$ \citep[e.g.,][]{manara_x-shooter_2020, testi_protoplanetary_2022}. Assuming that all the matter in the disk is eventually accreted by the star gives $\Delta M/M_0 \approx 10^{-3}-10^{-1}$. In practice, the accreted mass will be smaller, due to photoevaporation \citep{somigliana_effects_2020,winter_external_2022} or magnetized winds \citep{lesur_magnetohydrodynamics_2021, turpin_orbital_2024}.
Arguably, the most studied Class II binary system interacting with a circumbinary disk is GG Tau \citep{dutrey_images_1994, guilloteau_gg_1999, scaife_long-wavelength_2013, andrews_resolved_2014, aly_2018}, but see also GM Au \citep{dutrey_co_1998}, TWA 3A \citep{tofflemire_pulsed_2017} the Herbig stars HD 142527 \citep{lacour_m-dwarf_2016, price_2018} and V892 Tau \citep{long_2021}.

Having revised the disk properties of YSOs, we conclude the following:
In the earliest phases of star formation, circumbinary disks are thick, self-gravitating or fragmenting, and result in high accretion rates ($>10^{-6} \si{\Msun\per\year}$) onto their YSOs. They later transition to a stable, thin configuration with a much lower accretion rate ($<10^{-7} \si{\Msun\per\year}$).
The amount of mass assembled in each phase, and the relative importance of steady versus bursty accretion is not yet completely understood.
However, as long as the final few percent of the mass is accreted from a stable thin circumbinary disk, during the Class I or Class II phases, it should be possible to discern the signatures of an interaction with the disk predicted in this work.
To our knowledge, such features in the main-sequence star populations have not been reported in the literature \citep[e.g.,][]{tokovinin_eccentricity_2020, hwang_eccentricity_2022, andrew_binary_2022, gaia_collaboration_gaia_2023}.

However, there exist some plausible explanations for the lack of observational evidence for our predicted signatures.
In the first place, the mass accreted from a thin, stable disk during star formation could account for less than $\sim$1\% of the mass of the binary. In this case, the effect of the interaction with the disk would be negligible (see Figs.~\ref{fig:e_q_correlation} and \ref{fig:evolve2_comparison}).
In the second place, some physical processes currently not included in our model (such as outflows, radiation transport, or dynamical interactions with other stars) may play an important role in the orbital evolution of binary YSOs.
In the third place, the evolution might diverge from our predictions if the assumption of a relaxed disk becomes invalid (see Sect.~\ref{sec:hydro_description}). Our approach assumes that the orbital parameters evolve on a longer timescale than the viscous time $t_{\rm visc}(r)~=~ 2/3~r^2/\nu(r)$ \citep[e.g.,][]{haiman_population_2009}. For the inner region $r=2a$, which dominates the forces, $t_{\rm visc} \approx 300 P_{\rm{orb}}$. We estimate the characteristic timescale for the change of orbital parameters as
\begin{equation}
\begin{split}
    t_X \equiv& \frac{1}{X'} \frac{M}{\dot M} = 1000 P_{\rm{orb}} \frac{1}{X'} \left(\frac{a}{\SI{100}{\au}}\right)^{-3/2}\\ &\left(\frac{M}{\si{\Msun}}\right)^{3/2} \left(\frac{\dot M}{10^{-6}\,\si{\Msun\per\year}}\right)^{-1},
\end{split}
\end{equation}
where $X$ stands for $a$, $e$ or $q$ and $X'$ is defined as in Equation \ref{eq:def_derivatives} and can be as high as $\sim 8$ (see Fig.~\ref{fig:edot_qdot}). Hence, in systems with a wide orbit and a high accretion rate, $t_X$ can be comparable or even shorter than $t_{\rm visc}$. This violates the assumption that the disk is relaxed, especially for wide binaries, where tidal effects are negligible. Furthermore, the hydrodynamic simulations used herein fix the binary orbit and do not account for the back-reaction from the surrounding gas \citep[see e.g.,][for discussion]{franchini_importance_2023, tiede_disk-induced_2024}
In the fourth place, the assumption that the size of the binary components is much smaller than the orbit may become invalid for short-period systems and when the effective accretion radius of the stars becomes large (e.g., because of the presence of a magnetosphere).
A fifth caveat concerns the treatment of viscosity. Magnetorotational instability, which is generally believed to source the Reynolds stresses that redistribute angular momentum in many disks, is significantly suppressed in cool proto-stellar disks \citep{king_accretion_2007}. Measurements of viscosity for Class II protostars indicate $3 \times 10^{-4} \leq \alpha \leq 3 \times 10^{-3}$ \citep{rosotti_empirical_2023}, much lower than the value $\alpha = 0.1$ employed in this work. A final possibility is that only a fraction of the binary systems undergo stable, thin, disk accretion. In this case, the observational signatures would not immediately stand out in the statistics of a general population. A similar idea was brought forward by \citet{el-badry_2019}, who proposed that if a subset of the binaries accreted a significant fraction of their mass from a circumbinary disk, they would appear as a population of equal mass "twin" binaries and explain the excess of $q\approx1$ systems in the observed distributions \citep[see also the discussion in][]{duffell_circumbinary_2020}.

Before concluding the discussion about the astrophysical implications of our models to binary star formation, we also mention that binary-disk interactions have been hypothesized as a possible mechanism to explain the observed increase of radial velocity dispersion in young clusters as they age \citep{sana_dearth_2017, ramirez-tannus_massive_2017,ramirez-tannus_young_2020, ramirez-tannus_relation_2021}.
However, to explain the considerable decrease in the minimum period, from \SI{3500}{\day} in the young cluster M17 to \SI{1.4}{\day} in the older clusters, a binary with an initial total mass of $M=\SI{20}{\Msun}$ should decrease their separation by more than two orders of magnitude. 
Based on our current findings, achieving such a significant reduction in separation from a stable accretion scenario would require the system to increase its mass nearly tenfold, while the mass available in the disk is typically less than $10^{-3}\,\si{\Msun}$ \citep[e.g.,][]{backs_massive_2023}.

\subsection{Mass transferring stars} \label{sec:application_interacting_stars}
Circumbinary disks can also be formed during nonconservative mass transfer episodes.
These circumbinary disks have been studied in the context of cataclysmic variables \citep{spruit_circumbinary_2001, taam_evolution_2001, dubus_structure_2002} and they have been observed around X-ray binaries \citep{blundell_ss_2008}.
We refrain from giving quantitative statements regarding the orbital evolution in these cases, since the gas does not accrete on the central binary, but emanates from it. 
Given this crucial difference, our models are not directly applicable.
However, this scenario is worth studying with detailed hydrodynamic models, as it has important consequences on the evolution of the binary and, for example, on the merger rate of gravitational-wave sources \citep[e.g.,][]{van_son_no_2022-1}.

A further occurrence of circumbinary disks is found during stable mass transfer in a triple stellar system, where in some cases the mass stream from the outer star can form a circumbinary disk around the inner binary \citep{de_vries_evolution_2014, leigh_mergers_2020, di_stefano_mass_2020, dorozsmai_stellar_2024}. There is currently no direct observational counterpart of this process, but it is expected to be a natural stage of triple star evolution \citep{tauris_formation_2014, hamers_statistical_2022}. The effect of the disk on the orbit during triple mass transfer has often been neglected in past population synthesis studies. 
We believe that \texttt{spindler} will help to improve on this aspect, by providing a simple interface to access the angular momentum and orbital energy exchanged with the disk.

\subsection{Post--common-envelope systems} \label{sec:application_post_ce}
A scenario wherein the interaction with a circumbinary disk may play a crucial role is during the post--common-envelope phase. Following the initial dynamical phase of plunge-in, 3D simulations show that the inspiral process may stall even if the envelope has not been completely ejected. The stalling is caused by the formation of a low gas-density region at the envelope's center, where friction is reduced \citep{passy_simulating_2012, ricker_amr_2012, ricker_common_2019, moreno_3d_2022, lau_common_2022-1, lau_common_2022}.
Subsequently, the bound part of the remaining envelope is expected to cool and form a centrifugally supported disk, although the process takes too long to be captured in hydrodynamic simulations of the post--common-envelope phase \citep{reichardt_extending_2019,gagnier_post-dynamical_2023, gagnier_post-dynamical_2024}.

The interaction of the binary with the disk has been proposed as a mechanism to further reduce the orbital separation \citep{kashi_circumbinary_2011, tuna_long-term_2023, wei_evolution_2023}, which could potentially lead to a late merger.
Consequently, the final state of the system could be meaningfully influenced by the interaction with a disk, carrying crucial implications for double compact object mergers and gravitational-wave detection rates \citep[e.g.,][]{belczynski_rarity_2007,stevenson_formation_2017}.

Although the eccentricity of the binary before the common envelope can be arbitrarily high \citep[e.g.,][]{2020PASA...37...38V}, it is expected that when the plunge-in stalls the orbit will have become nearly circular, with only a small residual eccentricity $e<0.2$ \citep{trani_revisiting_2022}. 
Indeed, observed post--common-envelope candidates (identified as short period binaries with at least one hydrogen-depleted component) appear to favor small eccentricities \citep{kruckow_catalog_2021}.
Such post--common-envelope systems most prominently accentuate the model differences discussed in Sect.~\ref{sec:comparing_results}.
Namely, our fiducial model (\citetalias{siwek_orbital_2023}; Section~\ref{sec:results}) predicts that the orbit of accreting post--common-envelope systems will shrink; but only after their eccentricity has been excited to $e\approx 0.5$.
In contrast, the models based on \citetalias{zrake_equilibrium_2021} and \citetalias{dorazio_orbital_2021} project that systems with low residual eccentricity after the plunge-in will circularize; and that the disk will widen their orbital separations instead of shrinking them. 
The latter, we note, is more consistent with the observed small eccentricities of post--common-envelope systems.
Therefore, of the models considered, none provide a mechanism to reduce the orbital separation while maintaining consistency with the observed low eccentricities of post--common-envelope systems.

The previous considerations hinge on the assumption that the mass that can be accreted from a post--common-envelope circumbinary disk is sufficient to induce orbital change.
The amount of residual envelope mass after the plunge-in phase, however, can vary greatly.
For example, \citet{passy_simulating_2012} finds that a $\approx \SI{1}{\Msun}$ donor can retain more than 90\% of the envelope.
In contrast, \citet{lau_common_2022} finds that the whole envelope of a $\approx \SI{12}{\Msun}$ donor can be ejected.
In the case there is enough mass left to form a disk, \citet{tuna_long-term_2023} show that, for their chosen configuration, up to $\Delta M/M_0 = 10\%$ is accreted. 
This would be sufficient to pump (damp) binary eccentricity to the equilibrium band (circularity), but would not be adequate to significantly alter the orbital semi-major axis.

\subsection{Post-AGB binaries} \label{sec:application_post_agb}

Circumbinary disks are also thought to form from the asymmetric outflows of AGB stars in binary systems \citep{mohamed_mass_2012, saladino_eccentric_2019}, and are commonly observed around systems containing a post-AGB star \citep{de_ruyter_keplerian_2006, van_winckel_post-agb_2009, bujarrabal_extended_2013-1, gallardo_cava_keplerian_2021, kluska_population_2022, corporaal_transition_2023}.
Post-AGB systems are usually composed of a main-sequence star and a compact white dwarf or subdwarf companion, and have periods ranging from hundreds to thousands of days.
A long-standing problem resides in the eccentricity of these systems, which for $P_{\rm{orb}}<\SI{2000}{\day}$ should have been tidally circularized during the AGB phase  \citep{pols_can_2003, izzard_white-dwarf_2010}, however, they are found to have eccentricities up to $e\approx 0.6$ \citep{jorissen_insights_1998, oomen_orbital_2018}.
Eccentricity pumping via a circumbinary disk has been invoked to explain this phenomenon \citep{dermine_eccentricity-pumping_2013, antoniadis_formation_2014, vos_testing_2015, izzard_circumbinary_2023}, although \citet{rafikov_eccentricity_2016} argues against this possibility.

Moreover, \citet{oomen_orbital_2018} measure a bimodal distribution of eccentricities in post-AGB binaries with peaks around $e=0$ and $e=0.25$ (see also the discussion in \citetalias{zrake_equilibrium_2021}) which, if driven by interaction with a circumbinary disk, would suggest that near-circular accreting binaries tend to remain circular as in \citetalias{zrake_equilibrium_2021} and \citetalias{dorazio_orbital_2021}. 

Our findings show that the mass accreted from the disk needs to be at least 5-10\% the initial binary mass in order to drive significant orbit evolution, and the low masses ($10^{-3}\,-\,10^{-2}\, M_\odot$) contained in post-AGB disks \citep{bujarrabal_extended_2013-1} are likely insufficient. Moreover, the effects of disk winds and photoevaporation have not been included in the hydrodynamic models, but would likely amplify this issue.

\subsection{Supermassive black hole binaries} \label{sec:application_SMBH}

Supermassive black hole binaries are expected to form as a result of galaxy mergers \citep{begelman_1980}. 
If gas is present in the nuclear region, it can settle into a rotationally supported circumbinary disk \citep{barnes_formation_2002}.
The gas supply feeding the disk is self-regulated by the black hole accretion feedback, causing accretion to happen in short $\approx 10^5 \,\si{\year}$ bursts, during which the black holes can accrete mass close to the Eddington limit \citep{schawinski_active_2015}. 
Bursts are spaced out by quiescent phases, where the accretion luminosity can drop to $10^{-5}~L_{\rm{Edd}}$ \citep{novak_feedback_2011}. Active and quiescent phases can alternate $100$ to $10\,000$ times, comprising an Active Galactic Nucleus (AGN) phase lasting $10^7-10^9~\si{\year}$ \citep{martini_qso_2004, schawinski_active_2015}. 
\citet{siwek_effect_2020} showed that, within a realistic cosmological framework, SMBH binaries with $M<10^8\,\si{\Msun}$ can double or triple their initial mass via disk accretion before they merge, whereas, as the initial mass of the binaries increases, their ability to significantly increase their mass diminishes. These results depend on the disk torque model employed (in this case based on the analytic work of \citealt{haiman_population_2009}) and would likely vary when using the prescriptions derived in this work. Nonetheless, disk accretion can significantly increase the mass of SMBH binaries, rendering them an ideal scenario for studying the long-term evolution of the orbit with our formalism.

However, the simulations underlying this work assume a disk aspect ratio of $h/r=0.1$, which is notably higher than the typical values of $h/r \approx 10^{-2}-10^{-3}$ expected for supermassive binaries accreting from stable, thin disks (see Appendix~\ref{app:estimate_of_aspect_ratio}); particularly relevant for LISA progenitors ($10^5-10^7 \si{\Msun}$). Moreover, observations of AGN variability point to a viscosity $0.01 \leq \alpha \leq 0.03$ \citep{starling_constraints_2004, king_accretion_2007}, up to an order of magnitude lower than the value we assume.
Simulating the low viscosity and/or low aspect ratio regime poses numerical challenges, but it may present with important morphological and evolutionary differences from the $h/r=0.1$, $\alpha=0.1$ case \citep{tiede_gas-driven_2020,dittmann_survey_2022,dittmann_evolution_2024}.
Therefore, while our results are subject to change as one decreases viscosity and disk scale height, if a similar eccentricity attractor exists within this regime, we would expect a large population of supermassive binaries to maintain a relatively large eccentricity $e\sim0.5$, which could be manifest in electromagnetic searches.
This eccentricity will decrease once gravitational-wave emission starts dominating the orbital evolution, but a detectable nonzero residual eccentricity should still be present when entering the LISA band \citep{cuadra_2009,roedig_evolution_2012,zrake_equilibrium_2021, garg_minimum_2024}.
It is important to mention that we are only considering comparable-mass binaries ($q>0.1$) in our discussion. The dynamics of SMBHs with much lower mass ratios are more similar to protoplanetary rather than stellar-binary systems \cite[e.g.,][]{armitage_accretion_2002}.

Black hole binaries detectable through Pulsar Timing Arrays are more massive (above $10^8-10^9\,\si{\Msun}$) and have wider separations, and as such often lie outside the self-gravitating radius in standard $\alpha$-disk-like solutions \citep{haiman_population_2009}. 
Thus, the models based on such solutions here presented (and Equation~\ref{eq:aspect_ratio_smbh}) are strictly speaking not directly applicable. 
However, alternative solutions can preserve the disk's stability in such scenarios \citep{sirko_goodman:2003, gilbaum_stone:2022}, and gravitational stresses in an unstable disk may still be characterizable with an $\alpha$-prescription \citep{rice_2005}.
Moreover, \citet{roedig_limiting_2011} found that self-gravitating (but nonfragmenting) disks may display an equilibrium eccentricity around $e\approx 0.5$, similar to the ones underpinning the models in this work.
Such accretion around SMBH binaries in the PTA band could influence the power spectrum of the gravitational-wave background 
\citep{agazie_nanograv_2023-2, epta_collaboration_second_2023, reardon_parkes_2023, xu_cpta_2023}. For example, differential accretion can raise the median mass ratio of coalescing black holes and increase the power spectrum amplitude \citep{siwek_effect_2020}.
The power spectrum turnover at the lowest frequencies \citep[which was tentatively observed][]{agazie_nanograv_2023, epta_collaboration_second_2024} 
could also be accounted for with a population of eccentric binaries that shift gravitational wave emission to higher harmonic frequencies (although eccentricities higher than $0.5$ are required to significantly alter the spectral shape)
\citep{sesana_gravitational_2009,  sesana_insights_2013, chen_efficient_2017}
or with environment induced orbital decay \citep{kocsis_gas-driven_2011, agazie_nanograv_2023}; 
each of which would be consistent with sustained periods of circumbinary accretion.

\section{Conclusion} \label{sec:conclusion}

In this study, we explore the long-term orbital evolution of binary systems interacting with circumbinary disks. Our model is based on the result of the hydrodynamic simulations by \citetalias{siwek_orbital_2023} that fully resolve the cavity region and the circumstellar disks, self-consistently accounting for gravitational and accretion forces. Our findings can be summarized as follows:

\begin{itemize}
    \item To alter significantly the orbital separation, the system needs to accrete a mass comparable to the total mass of the binary (\ref{sec:trajectories} and \ref{sec:orbital_decay})
    \item If the initial eccentricity exceeds about $0.1$, two consistent trends emerge (\ref{sec:trajectories}):
    \begin{itemize}
        \item After accreting $\approx 10\%$ of the initial mass, the eccentricity will stabilize at an equilibrium value $e\approx0.5$ that depends weakly on the mass ratio.
        \item After accreting a mass comparable to the mass of the binary, the mass ratio will approach unity, and the orbital separation will shrink appreciably.
    \end{itemize}
    \item For initial eccentricities below approximately $0.1$, our three models yield contrasting predictions. The model based on \citetalias{siwek_orbital_2023} suggests that the eccentricity will be excited to $e\approx 0.5$, and then the orbit will shrink, while the models based on  \citetalias{zrake_equilibrium_2021} and \citetalias{dorazio_orbital_2021} predict that the orbit will circularize and then widen. The difference between the two predictions hinges on the different signs of the eccentricity derivative $e'$ when $e<0.1$ (\ref{sec:behaviour_small_e}).
\end{itemize}

As a consequence of the above findings, a population of binaries that are interacting (or have recently interacted) with a thin, stable circumbinary disk will show some clear signatures in the orbital parameters distributions:
\begin{itemize}
    \item \citetalias{siwek_orbital_2023}: a correlation between mass ratio and eccentricity (\ref{sec:population}).
    \item \citetalias{zrake_equilibrium_2021} and \citetalias{dorazio_orbital_2021}: a gap in the eccentricity distribution at $e \approx 0.1$ (\ref{sec:population_comparison}).
\end{itemize}

We discuss the applicability of these results to various astrophysical scenarios, namely binary star formation (\ref{sec:application_star_formation}), interacting stars (\ref{sec:application_interacting_stars}), post--common-envelope systems (\ref{sec:application_post_ce}), post-AGB binaries (\ref{sec:application_post_agb}) and SMBH binaries (\ref{sec:application_SMBH}).
We conclude that in almost all the considered scenarios, the systems are unable to accrete from the disk enough mass to significantly change the orbital separation. However, the interaction with the disk may still leave the above-mentioned signatures on the eccentricity distribution.
A notable exception arises with SMBH binaries, as they can substantially increase mass through disk accretion. Consequently, the interaction with the disk will impact also the mass ratio and semi-major axis distribution, with implications for LISA and PTA observations.
We note that the generality of these results is limited by our assumptions. We base our work on a grid of 2D simulations, while 3D effects are likely  important \citep{duffell_santa_2024}. We assume a fixed disk aspect ratio and viscosity prescription, while disks in nature can span a wide range of these parameters (see Fig.~\ref{fig:astro_disks}). Similarly, we ignore the effect of self-gravity, disk winds and disk-orbit inclination, which can all become important for specific applications. The vast parameter space that ensues from these variations has not been systematically explored and will require further investigation before these effects can be included in a formalism like the one we presented here.

\begin{acknowledgements}
RV thanks Norbert Langer, Shazrene Mohamed, Carlos Badenes, and all the participants of the MPA-Kavli Summer Program 2023 for the useful discussions and suggestions. A special thanks to Paul Ricker, for providing valuable insight on the (post-)common envelope phase, to Alice Somigliana, for the eye-opening discussion on protostellar disks, and to Deepika Bollimpalli, for answering far too many questions about AGN disks. RV and SdM are also grateful to  Giuseppe Lodato and Alessia Franchini for discussions the numerical approach and 3D effects. The authors also thank the referee for a thorough and constructive report.
DD and CT acknowledge support from the Danish Independent Research Fund through Sapere Aude Starting Grant No. 121587.
JC acknowledges support from ANID, Millenium Nuclei NCN19\_171 (NPF) and NCN2023{\_}002 (TITANs). This research was supported in part by grant NSF PHY-1748958 to the Kavli Institute for Theoretical Physics (KITP). This project was part of the Kavli Summer Program in Astrophysics 2023, hosted at the Max Planck Institute for Astrophysics. We thank the Kavli Foundation and the MPA for their support.
\end{acknowledgements}

\textit{Software:} NumPy \citep{harris_array_2020}, SciPy \citep{virtanen_scipy_2020}, Matplotlib \citep{hunter_matplotlib_2007}.

\bibliographystyle{aa}
\bibliography{main}

\begin{appendix}

\section{Estimate of the disk aspect ratio}
\label{app:estimate_of_aspect_ratio}
We can estimate the disk aspect ratio $h/r$ of a thin accretion disk via equation 2.19 of \citet{shakura_black_1973}, which, for parameters typical of an accreting stellar binary gives

\begin{equation}
\begin{split}
\label{eq:aspect_ratio_star}
    h/r \approx \; 0.12 &\left(\frac{M}{\si{\Msun}}\right)^{-3/8} \left(\frac{r}{\si{\au}}\right)^{1/8}
    \left(\frac{\alpha}{0.1}\right)^{-1/10} \left(\frac{\dot M}{10^{-6} \, \si{\Msun\per\year}}\right)^{3/20},
\end{split}
\end{equation}

where $M$ is the mass of the central object (in this context, the total mass of the binary), $r$ is the radial coordinate of the disk, to be taken as comparable to the size of the circumbinary cavity, $\alpha$ is the Shakura-Sunyaev viscosity parameter and $\dot M$ is the mass accretion rate.

When rewriting Equation~\ref{eq:aspect_ratio_star} for parameters typical of SMBHs, a significantly lower aspect ratio is obtained.

\begin{equation}
\begin{split}
\label{eq:aspect_ratio_smbh}
    h/r \approx \; 6.2 \times 10^{-3} &\left(\frac{M}{10^6 \, \si{\Msun}}\right)^{-1/10} \left(\frac{r}{10^4 \, r_g}\right)^{1/8} \\
     &\left(\frac{\alpha}{0.1}\right)^{-1/10} \left(\frac{L}{L_{\rm Edd}}\right)^{3/20} \left(\frac{\eta}{0.06}\right)^{-3/20},
\end{split}
\end{equation}

where $r_g=\frac{2GM}{c^2}$ is the Schwarzschild radius, $L=\eta \dot M c^2$ is the disk luminosity, $\eta$ is the gravitational energy release efficiency, and $L_{\rm Edd}$ is the Eddington luminosity.

\section{Additional figures}
\label{app:additional_figures}

\begin{figure}
    \centering
    \includegraphics[width=0.5\textwidth]{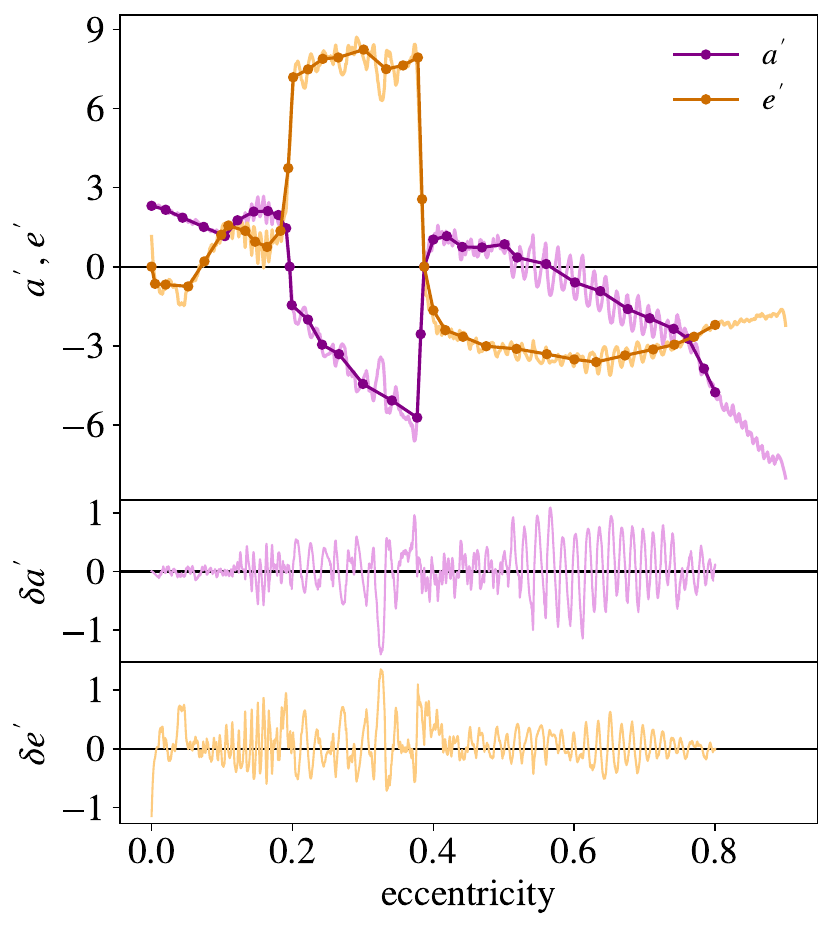}
    \caption{Interpolation of the results of \citetalias{dorazio_orbital_2021}. In the top panel, a comparison between the values of $a'$ and $e'$ computed in \citetalias{dorazio_orbital_2021} (lighter solid lines) and the interpolating function used in this work (darker solid lines with dots). The bottom panels show the residuals.}
    \label{fig:DD_derivatives}
\end{figure}

We describe two additional figures.
Figure~\ref{fig:DD_derivatives} shows our interpolation of the results of \citetalias{dorazio_orbital_2021}. As discussed in Sect.~\ref{sec:interpolation}, the functions $a'$ and $e'$ by \citetalias{dorazio_orbital_2021} display ambiguous oscillations. 
They are likely caused by the precession of the disk cavity, which happens on a timescale of a few hundred orbits and therefore is not completely smoothed out by the $\approx 444$ orbits window of the filter employed by \citetalias{dorazio_orbital_2021}. 
We perform a discrete resampling of the curves, selecting the points in such a way as to smooth out the oscillations, while at the same time preserving the underlying trend and the points where the curves change sign. In the bottom panels, we show the residuals of the linear interpolation.

Figure~\ref{fig:derivatives_q1} compares the results of the hydrodynamic simulations by \citetalias{siwek_orbital_2023}, \citetalias{zrake_equilibrium_2021} and \citetalias{dorazio_orbital_2021} for $q=1$.
The general trend is similar among the three sets of simulations, the most relevant differences for the long-term evolution of the orbit are the discrepancy in the sign of $e'$ for $e<0.1$ and the different location of the equilibrium eccentricity.
The former difference implies that the predictions of the three models diverge for systems that start with $e<0.1$ (Section~\ref{sec:behaviour_small_e}). The latter determines a different speed in the semi-major axis evolution once systems have reached the equilibrium eccentricity (Section~\ref{sec:orbital_decay}).

\begin{figure}
    \centering
    \includegraphics[width=0.5\textwidth]{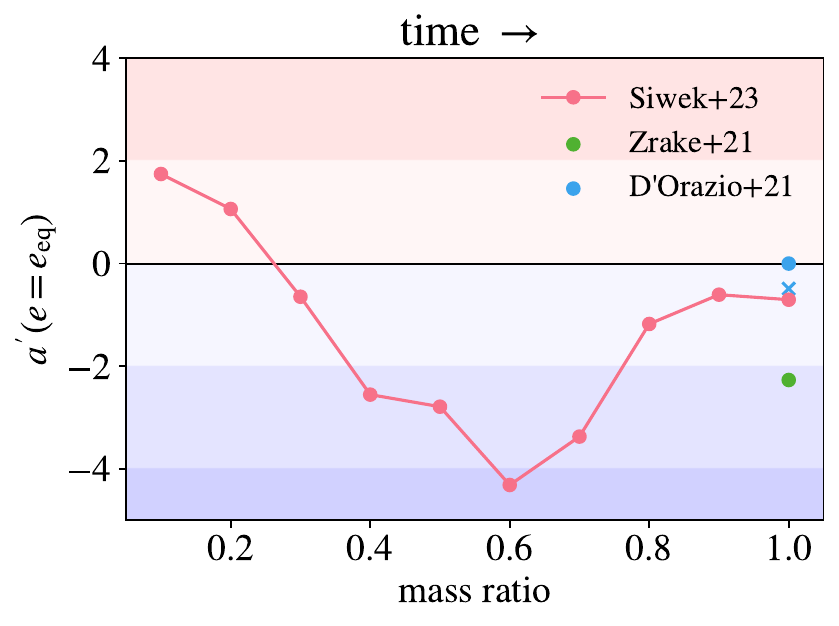}
    \caption{Rate of separation change $a'$ computed at the equilibrium eccentricity, as a function of the mass ratio. We compare the models based on \citetalias{siwek_orbital_2023} (red), \citetalias{zrake_equilibrium_2021} (green) and \citetalias{dorazio_orbital_2021} (blue). The blue cross indicates the value obtained by \citetalias{dorazio_orbital_2021} when considering small oscillations that a realistic system can have around the equilibrium eccentricity. Time moves from left to right as the binary accretes mass and the mass ratio approaches unity.}
    \label{fig:eq_ecc_adot}
\end{figure}

\begin{figure}
    \centering
    \includegraphics[width=0.5\textwidth]{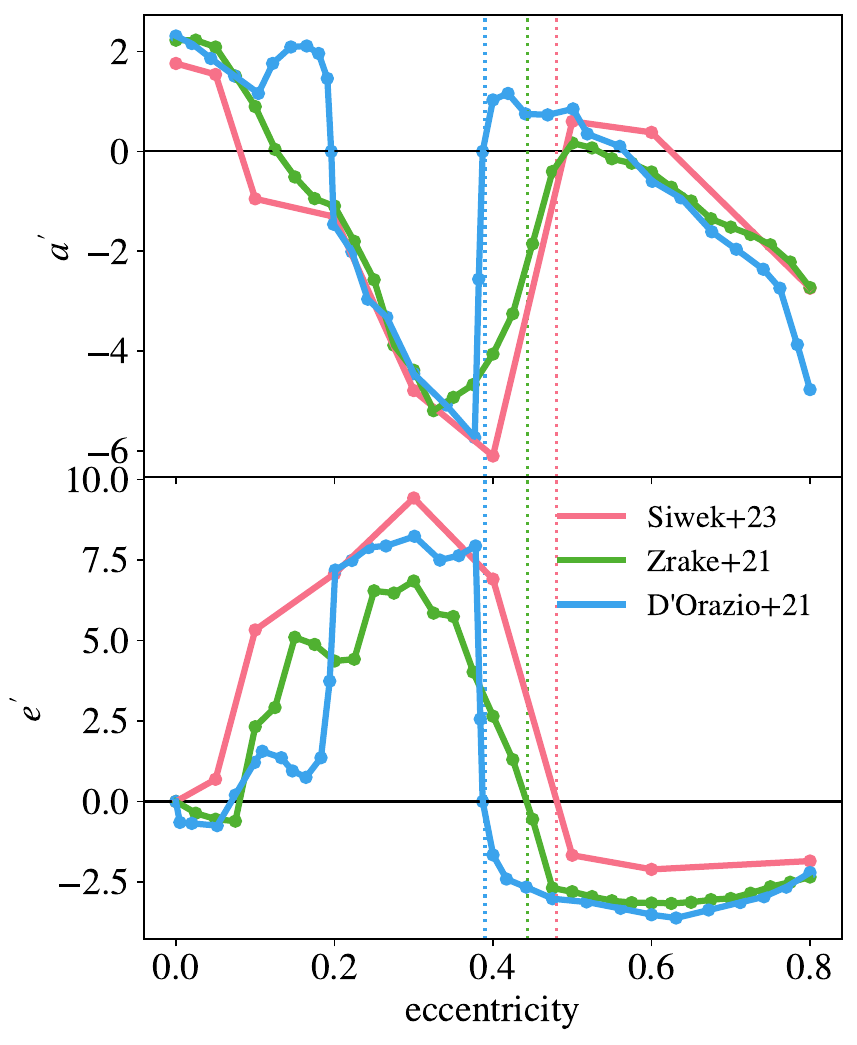}
    \caption{Results from the literature. Comparison of hydrodynamic simulations of an equal-mass binary and different eccentricities in a circumbinary disk from \citetalias{siwek_orbital_2023} (red), \citetalias{zrake_equilibrium_2021} (green) and \citetalias{dorazio_orbital_2021} (blue). The dots correspond to the simulations and the solid lines correspond to our interpolation. The top panel shows the value of the derivative of the semi-major axis $a'$, while the bottom panel shows the derivative of the eccentricity $e'$. Vertical dotted lines mark the value of the equilibrium eccentricity, where $e' = 0$.}
    \label{fig:derivatives_q1}
\end{figure}

Figure~\ref{fig:eq_ecc_adot} shows how fast the evolution of the semi-major axis is for systems at the equilibrium eccentricity. Systems with small $q<0.2$ widen their orbit, while for larger mass ratios the orbit always shrinks.
The equilibrium eccentricity is reached after binaries accrete only about $10\%$ of their initial mass (Section~\ref{sec:trajectories}) and, after reaching it, they remain close to the equilibrium as the mass ratio steadily increases. Therefore, the mass ratio axis can be also interpreted as a time evolution axis.
In this sense, a binary system that stars with $q<0.2$ will initially increase its orbital separation, then it will transition to orbital shrinking as the mass ratio becomes larger than $\approx 0.3$. The fastest orbital shrinking ($a'\approx -4$) takes place when the binary passes from $q\approx 0.6$. In the end, when the system reaches the attractor at $q=1$, the rate of separation change settles to $a' \approx -0.7$.
We also show the corresponding values for \citetalias{zrake_equilibrium_2021} and \citetalias{dorazio_orbital_2021} at $q=1$.

\FloatBarrier

\onecolumn
\section{Tables of derivatives}
\label{app:tables_of_derivatives}
We report here for convenience the tables of derivatives used for the interpolation, as described in Sect.~\ref{sec:methods}.

\begin{table}[!ht]
    \caption{Values of the semi-major axis derivative $a'$ from \citetalias{siwek_orbital_2023}.}
    \begin{tabular}{|c|ccccccccc|}
    \hline
        \diagbox[innerleftsep=5pt,innerrightsep=5pt,width=25pt, height=15pt]{$q$}{$e$} & 0.0 & 0.05 & 0.1 & 0.2 & 0.3 & 0.4 & 0.5 & 0.6 & 0.8 \\ \hline
        0.1 & -1.28 & -3.34 & -5.06 & 1.03 & 3.43 & 3.74 & 4.0 & 3.0 & -6.32 \\ 
        0.2 & -0.77 & -0.49 & -1.51 & -0.16 & 0.92 & 2.87 & 2.59 & -1.3 & -7.09 \\ 
        0.3 & 1.15 & 0.87 & -2.05 & -1.89 & -0.19 & -1.44 & -0.93 & -2.34 & -3.49 \\ 
        0.4 & 1.29 & 1.14 & -1.3 & -0.65 & -2.41 & -2.5 & -2.93 & -1.48 & -3.61 \\ 
        0.5 & 1.43 & 1.41 & -0.69 & -0.15 & -2.43 & -2.1 & -3.73 & -1.26 & -3.52 \\ 
        0.6 & 1.58 & 1.50 & -0.69 & -0.42 & -2.37 & -2.96 & -4.33 & -0.3 & -2.73 \\ 
        0.7 & 1.67 & 1.52 & -0.75 & -0.46 & -2.38 & -5.16 & -4.36 & 0.28 & -2.85 \\ 
        0.8 & 1.72 & 1.54 & -0.94 & -0.67 & -2.52 & -6.23 & -0.28 & 0.52 & -3.00 \\ 
        0.9 & 1.74 & 1.55 & -0.88 & -1.02 & -4.15 & -6.23 & 0.86 & 0.47 & -2.89 \\ 
        1.0 & 1.76 & 1.54 & -0.95 & -1.31 & -4.79 & -6.10 & 0.60 & 0.38 & -2.74 \\  \hline
    \end{tabular}
    \label{tab:adota_siwek}
\end{table}

\begin{table}[!ht]
    \caption{Values of the eccentricity derivative $e'$ from \citetalias{siwek_orbital_2023}.}
    \begin{tabular}{|c|ccccccccc|}
    \hline
        \diagbox[innerleftsep=5pt,innerrightsep=5pt,width=25pt, height=15pt]{$q$}{$e$}  & 0.0 & 0.05 & 0.1 & 0.2 & 0.3 & 0.4 & 0.5 & 0.6 & 0.8 \\ \hline
        0.1 & 0.0 & -0.47 & 1.55 & 0.78 & -1.84 & -4.15 & -4.78 & -5.95 & -7.7 \\ 
        0.2 & 0.0 & 0.02 & 1.32 & 2.14 & 0.16 & -2.02 & -3.96 & -4.62 & -5.47 \\ 
        0.3 & 0.0 & 0.13 & 3.73 & 5.59 & 0.23 & -0.40 & -2.73 & -3.95 & -3.46 \\ 
        0.4 & 0.0 & 0.54 & 4.29 & 3.50 & 2.52 & 0.23 & -1.64 & -2.81 & -2.61 \\ 
        0.5 & 0.0 & 0.77 & 4.33 & 3.75 & 3.38 & 1.33 & -1.82 & -2.37 & -2.15 \\ 
        0.6 & 0.0 & 1.00 & 4.73 & 4.90 & 4.52 & 3.33 & -0.04 & -2.20 & -1.96 \\ 
        0.7 & 0.0 & 0.92 & 4.88 & 5.48 & 5.26 & 5.60 & 0.58 & -2.14 & -1.86 \\ 
        0.8 & 0.0 & 0.81 & 5.28 & 5.95 & 5.97 & 6.48 & -1.15 & -2.08 & -1.70 \\ 
        0.9 & 0.0 & 0.75 & 5.16 & 6.60 & 8.33 & 7.02 & -1.83 & -2.12 & -1.69 \\ 
        1.0 & 0.0 & 0.69 & 5.33 & 7.07 & 9.43 & 6.91 & -1.67 & -2.11 & -1.85 \\ \hline
    \end{tabular}
    \label{tab:edot_siwek}
\end{table}

\begin{table}[!ht]
    \caption{Values of the relative accretion rate $\lambda$ from \citetalias{siwek_preferential_2023}.}
    \begin{tabular}{|c|ccccc|}
    \hline
        \diagbox[innerleftsep=5pt,innerrightsep=5pt,width=25pt, height=15pt]{$q$}{$e$} & 0.0 & 0.2 & 0.4 & 0.6 & 0.8 \\ \hline
        0.1 & 6.51 & 4.55 & 2.00 & 1.24 & 1.34 \\ 
        0.2 & 3.94 & 7.05 & 2.16 & 1.60 & 1.51 \\ 
        0.3 & 3.05 & 4.82 & 2.67 & 1.84 & 1.97 \\ 
        0.4 & 2.28 & 3.43 & 3.19 & 3.30 & 1.30 \\ 
        0.5 & 2.03 & 3.41 & 4.79 & 3.33 & 9.26 \\ 
        0.6 & 1.86 & 3.21 & 4.08 & 1.88 & 8.60 \\ 
        0.7 & 1.63 & 2.96 & 1.97 & 1.35 & 8.56 \\ 
        0.8 & 1.40 & 2.51 & 1.64 & 1.13 & 8.97 \\ 
        0.9 & 1.17 & 2.22 & 1.29 & 1.09 & 9.91 \\ 
        1.0 & 1.00 & 1.00 & 1.00 & 1.00 & 1.00 \\ \hline
    \end{tabular}
    \label{tab:lambda_siwek}
\end{table}

\begin{table}[!ht]
    \caption{Values of the semi-major axis derivative $a'$ and eccentricity derivative $e'$ derived from the results of \citetalias{zrake_equilibrium_2021} (columns 1-3) and \citetalias{dorazio_orbital_2021} (columns 4-7).}
    \begin{tabular}{|ccc|cc|cc|}
    \hline
    \multicolumn{3}{|c}{\citetalias{zrake_equilibrium_2021}} & \multicolumn{4}{|c|}{\citetalias{dorazio_orbital_2021}} \\
    \hline
    $e$   & $a'$  & $e'$  & $e$   & $a'$  & $e$   & $e'$  \\
    \hline
    0     & 2.23  & 0     & 0     & 2.31  & 0     & 0     \\
    0.025 & 2.23  & -0.36 & 0.02  & 2.16  & 0.005 & -0.65 \\
    0.05  & 2.09  & -0.55 & 0.044 & 1.86  & 0.02  & -0.68 \\
    0.075 & 1.51  & -0.61 & 0.074 & 1.51  & 0.052 & -0.75 \\
    0.1   & 0.9   & 2.32  & 0.104 & 1.16  & 0.075 & 0.2   \\
    0.125 & 0.04  & 2.92  & 0.122 & 1.76  & 0.099 & 1.21  \\
    0.15  & -0.51 & 5.1   & 0.145 & 2.09  & 0.109 & 1.56  \\
    0.175 & -0.95 & 4.88  & 0.165 & 2.11  & 0.133 & 1.36  \\
    0.2   & -1.09 & 4.36  & 0.18  & 1.96  & 0.147 & 0.95  \\
    0.225 & -1.8  & 4.42  & 0.191 & 1.46  & 0.164 & 0.75  \\
    0.25  & -2.57 & 6.55  & 0.196 & 0     & 0.183 & 1.36  \\
    0.275 & -3.88 & 6.48  & 0.199 & -1.46 & 0.194 & 3.74  \\
    0.3   & -4.38 & 6.85  & 0.222 & -2.01 & 0.201 & 7.19  \\
    0.325 & -5.19 & 5.85  & 0.242 & -2.96 & 0.222 & 7.49  \\
    0.35  & -4.93 & 5.75  & 0.266 & -3.32 & 0.243 & 7.89  \\
    0.375 & -4.67 & 4.02  & 0.3   & -4.45 & 0.265 & 7.94  \\
    0.4   & -4.06 & 2.65  & 0.341 & -5.08 & 0.301 & 8.24  \\
    0.425 & -3.25 & 1.3   & 0.377 & -5.73 & 0.333 & 7.5   \\
    0.45  & -1.85 & -0.55 & 0.382 & -2.56 & 0.357 & 7.64  \\
    0.475 & -0.4  & -2.69 & 0.387 & 0     & 0.378 & 7.94  \\
    0.5   & 0.17  & -2.8  & 0.4   & 1.03  & 0.384 & 2.56  \\
    0.525 & 0.07  & -2.96 & 0.419 & 1.16  & 0.387 & 0     \\
    0.55  & -0.15 & -3.09 & 0.441 & 0.75  & 0.4   & -1.66 \\
    0.575 & -0.24 & -3.14 & 0.469 & 0.73  & 0.417 & -2.41 \\
    0.6   & -0.41 & -3.16 & 0.501 & 0.85  & 0.442 & -2.66 \\
    0.625 & -0.72 & -3.17 & 0.519 & 0.35  & 0.475 & -3.02 \\
    0.65  & -1    & -3.13 & 0.56  & 0.1   & 0.518 & -3.12 \\
    0.675 & -1.35 & -3.05 & 0.601 & -0.6  & 0.561 & -3.32 \\
    0.7   & -1.52 & -3.01 & 0.637 & -0.93 & 0.6   & -3.52 \\
    0.725 & -1.67 & -2.85 & 0.676 & -1.61 & 0.631 & -3.62 \\
    0.75  & -1.87 & -2.65 & 0.707 & -1.96 & 0.672 & -3.37 \\
    0.775 & -2.22 & -2.51 & 0.741 & -2.36 & 0.712 & -3.14 \\
    0.8   & -2.73 & -2.35 & 0.762 & -2.74 & 0.742 & -2.96 \\
          &       &       & 0.784 & -3.87 & 0.77  & -2.66 \\
          &       &       & 0.8   & -4.77 & 0.8   & -2.21 \\
    \hline
    \end{tabular}
    \label{tab:derivatives_Z21_DD21}
\end{table}

\end{appendix}
\end{document}